\documentclass[10pt,preprint]{emulateapj}
\usepackage[utf8]{inputenc}
\usepackage{amsmath}
\usepackage{longtable}
\usepackage{graphicx}
\usepackage{mathtools}

\DeclarePairedDelimiter\abs{\lvert}{\rvert}
\begin{document}
\submitted{Accepted for Publication in the Astronomical Journal}
\title{\textit{Gaia} Assorted Mass Binaries Long Excluded from SLoWPoKES (GAMBLES): Identifying Ultra-Wide Binary Pairs with Components of Diverse Mass}

\author{Ryan J. Oelkers\altaffilmark{1*}, Keivan G. Stassun\altaffilmark{1,2}, Saurav Dhital\altaffilmark{1}}
\altaffiltext{1}{Vanderbilt University, Department of Physics and Astronomy, Nashville, TN 37235}
\altaffiltext{2}{Fisk University, Department of Physics and Astronomy, Nashville, TN 37208}
\altaffiltext{*}{Corresponding author: ryan.j.oelkers@vanderbilt.edu}

\begin{abstract}
The formation and evolution of binary star systems still remain key questions in modern astronomy. Wide binary pairs (separations $>10^3$~AU) are particularly intriguing because their low binding energies make it difficult for the stars to stay gravitationally bound over extended timescales,and thus probe the dynamics of binary formation and dissolution. Our previous SLoWPoKES~I~\&~II catalogs provided the largest and most complete sample of wide binary pairs of low masses.  Here we present an extension of these catalogs to a broad range of stellar masses: the \textit{Gaia} Assorted Mass Binaries Long Excluded from SloWPoKES (GAMBLES), comprising 8,660 statistically significant wide pairs that we make available in a living online database. Within this catalog we identify a subset of 543 long-lived (dissipation timescale $>$1.5~Gyr) candidate binary pairs, of assorted mass, with typical separations between $10^3-10^{5.5}$~AU ($0.002-1.5$~pc), using the published distances and proper motions from the \textit{Tycho-Gaia} Astrometric Solution and Sloan Digital Sky Survey photometry. Each pair has \textit{at most} a false positive probability of 0.05; the total expectation is 2.44 false binaries in our sample. Among these, we find 22 systems with 3 components, 1 system with 4 components, and 15 pairs consisting of at least 1 possible red giant. We find the largest long-lived binary separation to be nearly 3.2~pc; even so, $>76\%$ of GAMBLES long-lived binaries have large binding energies and dissipation lifetimes longer than 1.5~Gyr. Finally, we find the distribution of binary separations is clearly bimodal, corroborating the finding from SloWPoKES and suggesting multiple pathways for the formation and dissipation of the widest binaries in the Galaxy.
\end{abstract}

\section{Introduction}

A wide range of astrophysical studies require an unqualified understanding of the fundamental properties of stars. Diverse topics such as planet, star \& galaxy formation, the initial stellar mass function, the distance scale and supernovae are just some examples of areas that benefit from realistic models of star formation and evolution \citep{Kennicutt:2012, Stassun:2014}. The study of binary star systems is particularly fruitful because binaries can provide direct measurements of the physical parameters of stars and provide insight into basic star formation processes and galactic dynamic evolution \citep{Shaya:2011}. 

The formation of binary systems is typically understood to occur during a coeval fragmentation of a giant molecular cloud. While this formation theory can explain the occurrence of binary star system with close separations \citep{Duquennoy:1991, Raghavan:2010}, the discovery of binary pairs with separations near the typical pre-stellar cores has strained typical formation models \citep{Elliott:2016}. Particularly interesting are very-wide binary pairs, ($>1$~pc separation) which provide a unique opportunity to study the formation, evolution and continued stability of binary pairs with separations too wide to be the result of the fragmentation of a molecular cloud. 

The creation and sustainability of wide-separation binaries is increasing difficult to explain with a synoptic formation model. The collapse of progenitor molecular clouds into wide-binary pairs has been hypothesized to be caused by the continued passing of numerous stars through the cloud, supernovae explosions, interactions with other giant molecular clouds or dynamical instabilities within the cloud itself \citep{Wasserman:1987, Palasi:2000, Reipurth:2012}. The wide separation of these systems, coupled with their immense orbital periods and the large timescale of dynamic interactions, prohibit a comprehensive study of a given system's orbital parameters through direct observation. Thus, wide binary formation and evolution theory is developed through the inference of the properties from the investigation of large-scale population statistics. This has proven to be a difficult task, given the dearth of confirmed wide-binary pairs with measured distances, proper motions and radial velocities \citep{Kouwenhoven:2010, Elliott:2016}. 

Recent technological advances have allowed a massive amount of astronomical data to be collected and reduced on practical timescales. Large time-series surveys, such as the Sloan Digital Sky Survey (hereafter, SDSS) \citep{York:2000} and the Two-Micron All Sky Survey (hereafter, 2MASS) \citep{Skrutskie:2006}, have provided some of the most exquisite astronomical data during the past two decades. The data products from these surveys have permitted massive data mining studies to discover underlying population characteristics and have answered many unresolved questions in Galactic evolution, extra-galactic formation and stellar theory.

The Sloan Low-mass Pairs of Kinematically Equivalent Stars (SLoWPoKES) I \& II catalogs (hereafter SLW I \& II) provided the largest, statistically significant sample of wide binary pairs to date \citep{Dhital:2010, Dhital:2015} and complimented similar studies of wide binary pairs \citep{Sesar:2008}. These catalogs were the result of a massive search of the SDSS dataset to identify and characterize low-mass, wide-binary pairs. These binaries included 1,342 low mass pairs which were identified with positions, distances and proper motions and an additional 105,537 pairs identified solely with positions and astrometry. While ground-breaking, these studies were limited by their use of distance-color relations rather than direct distance measurements.

Distance is quite possibly the most fundamental measurable parameter in astronomy but is almost never directly measured. The mission of the \textit{Gaia} space satellite is an ambitious survey with the goal of making a precise 3D map of nearly 100 billion stars over the course of the five year mission lifetime. Following in the footsteps of previous space based astrometric surveys, the satellite will return parallax measurements of stars and provide direct distance determinations \citep{vanLeeuwen:2007, Lindegren:2016}. The first data release provided parallax measurements for stars matched between the \textit{Gaia} data set and the Tycho-2 catalog \citep{Hog:2000, Lindegren:2016}. 

In this publication we identify systems previously excluded in the SLoWPoKES sample, high to medium mass wide binary pairs. We use the proper motions, parallaxes and positions from the first \textit{Gaia} data release, combined with the photometry, astrometry and proper motions from SDSS to identify binary pairs between multiple Tycho-2 stars and between Tycho-2 and SDSS point sources. We call this extension of the SLoWPoKES catalog the \textit{Gaia} Assorted Mass Binaries Long Excluded from SLoWPoKES (GAMBLES). 

The remainder of this paper is organized in the following way: \S~\ref{sec:data} describes the archival data; \S~\ref{sec:methods} describes our search for and confirmation of ultra-wide binary candidates through the use of a 5D Galactic Model; \S~\ref{sec:results} describes our results; \S~\ref{sec:discussion} is a discussion of our results; and \S~\ref{sec:conclusions} are our conclusions. 

\section{Survey Data \label{sec:data}}

\subsection{Gaia}

The \textit{Gaia} spacecraft was launched in December 2013 with a 5 year mission to survey the entire celestial sphere to collect trigonometric parallaxes for more than 100 billion stars. The telescope is expected to achieve an astrometric precision of 20$\mu$as or better for all stars with visual magnitudes brighter than $V\approx15$ \citep{vanLeeuwen:2007, Lindegren:2016}. 

The satellite has three main instruments: the astrometric field, blue \& red photometers and a radial velocity spectrometer. The astrometric field records white light (3300-10500 \AA, G-band) for the astrometric calculations. The blue (3300-6800 \AA) and red (6400-10500 \AA) photometers produce low-resolution spectra over 62 pixels. The radial velocity spectrometer has a resolution of 11,500 from 8470-8710 \AA\ to cover the calcium triplet and provide radial velocities for stars brighter than $V\approx 16$. The satellite is covered in 106 CCDs, making a nearly a Gigapixel camera. The spacecraft has two telescopes, each with a field of view of 0.25 sq. degrees, which rotate allowing for every source to be observed 72 times over the course of the 5 year mission \citep{vanLeeuwen:2007, Lindegren:2016}. 

The first data release from the survey provided positions, proper motions and parallaxes for $\sim2\times10^6$ \textit{Tycho-2} stars and $\sim9\times10^5$ \textit{Hipparcos} stars to create the \textit{Tycho-Gaia} Astrometric Solution catalog (hereafter, TGAS) \citep{ESA:1997, Hog:2000, vanLeeuwen:2007, Lindegren:2016}. This catalog is primarily composed of stars with $V<11.5$, contains typical uncertainties of 0.3 mas in the determined positions and parallaxes and 1 mas yr$^{-1}$ for the proper motions. Positions for more than $10^9$ stars were also released in \textit{Gaia}'s secondary data set. Future data releases are expected to include 5 parameter astrometric solutions for all stars determined to be in single star systems, followed by solutions for multiple star systems and culminating in a final data release in 2022 \citep{Lindegren:2016}.

We note that, at the time of this writing, the {\it Gaia\/} $\pi$ values potentially have systematic uncertainties that are not yet fully characterized but that could reach $\sim$300~$\mu$as\footnote{See \url{http://www.cosmos.esa.int/web/gaia/dr1}.}. Preliminary assessments suggest a global offset of $-0.25$~mas (where the negative sign indicates that the {\it Gaia\/} parallaxes are underestimated) for $\pi \gtrsim 1$~mas \citep{StassunGaiaEr:2016}, corroborating the {\it Gaia\/} claim, based on comparison to directly-measured distances to well-studied eclipsing binaries by \citet{Stassun:2016}. \citet{Gould:2016} similarly claim a systematic uncertainty of 0.12~mas.\citet{Casertano:2016} used a large sample of Cepheids to show that there is likely little to no systematic error in the {\it Gaia\/} parallaxes for $\pi \lesssim 1$~mas, but find evidence for an offset at larger $\pi$ consistent with \citet{StassunGaiaEr:2016}.
Thus the available evidence suggests that any systematic error in the {\it Gaia\/} parallaxes is likely to be small. Thus, for the purposes of this work, we use and propagate the reported {\it random} uncertainties on $\pi$ only, emphasizing that additional (or different) choices of $\pi$ uncertainties may be applied in the future, following the methodology laid out below.

We have chosen to incorporate the TGAS sample into the SLoWPoKES catalog for 3 reasons. First, the distances provided by the TGAS sample were directly measured through parallax and do not rely on the SDSS distance-color relations from previous SLoWPoKES work \citep{Dhital:2010,Dhital:2015}. Second, these objects were excluded from the original catalog because the faint limit of the TGAS sample, $V\approx12$, is saturated in the SDSS photometry. Finally, by including this catalog we can search for assorted mass binaries between a variety of spectral types, to probe a regime not yet studied in the SLoWPoKES sample. 

\subsection{SDSS}
The Sloan Digital Sky Survey (SDSS) is one of the most influential surveys in modern astronomy. The survey, currently on its thirteenth data release (DR-13) \citep{SDSS:2016}, marked the beginning of the big-data science era in astronomy and shepherded the future of many wide-field surveys such as the Large Synoptic Survey Telescope (LSST) \citep{Ivezic:2008} and the Transiting Exoplanet Survey Satellite (TESS) \citep{Ricker:2014}. The survey data has been collected by the 2.5~m telescope at the Apache Point Observatory since the year 2000. The telescope has a 120 mega-pixel camera with a field of view of 1.5 sq. degrees and conducts photometric observations in five optical broad bands, \textit{ugriz}, between 3000 and 10000 \AA  \citep{York:2000}. 

SDSS DR-13 contains more than 450 million unique objects over the entire 14,555 sq. degrees of the survey which spans the entire Northern sky and the Southern Galactic cap. The recalibration of the imaging data has allowed for a decreased systematic uncertainty of $<0.9\%$ in \textit{griz} \citep{SDSS:2016}. As in SLW I \& II, we used the Catalog Archive Server query tool (CasJobs) and the SciServer Compute python application to query within radial areas of each \textit{Tycho-2} star in TGAS, to select the sample of possible companions from SDSS. 

\section{Methods\label{sec:methods}}

\subsection{Identifying Wide Binary Candidates\label{sec:identify}}

The SLW-I\& II catalogs were the result of an extensive search for wide separation binaries in the SDSS dataset, which we extend in GAMBLES to include TGAS. While the final GAMBLES sample is a single catalog, it was formed in two distinct parts. The search of SDSS point sources for companions to TGAS stars and a search for wide binaries within the TGAS sample.

\subsubsection{Gaia Query\label{subsec:tgas}}

We queried TGAS using a cone search whose angular radius is a function of the distance to each TGAS star. Specifically, we defined our radial search to limit the maximum physical separation of the binary to be $\sim15$~pc. This evolving radius was adopted to avoid biasing the TGAS sample against discovering the widest binary pairs at close helio-centric distances. This sample is hereafter, TGAS-TGAS.

The TGAS stars have their distances measured directly using parallaxes from the \textit{Gaia} satellite \citep{Lindegren:2016}. We use the normal formula to convert from parallax to distance $d=\frac{1}{\pi}$, where $\pi$ is measured in arcseconds and $d$ is measured in parsecs. 

We elected to use the synthetic SDSS photometry provided by \citet{Pickles:2010} to estimate each spectral type since many stars in the TGAS sample never had SDSS \textit{ugriz} observations (at least not in a comprehensive, systematic way). These magnitudes were created by fitting the $B_T, V_T$ and 2MASS $JHK$ magnitudes to interpolate to the SDSS filters. We adopt the photometric uncertainties of $(0.2, 0.06, 0.04, 0.04, 0.05)$ in \textit{ugriz} respectively, based on the uncertainties provided in \citet{Pickles:2010}. We estimated the extinction to each star using the 3D dust map from \citet{Bovy:2016} and the TGAS distances. 

We then used the same relations described in \S~\ref{subsec:photrelations} to determine a photometric spectral type using the synthetic magnitudes and calculated extinction. Stars which did not fall into the color ranges provided by these relations were not assigned a spectral type but were included in our sample since their distances were based on parallax and not dependent on a color based relation.

Finally, we constrained our TGAS-TGAS sample by the following 5 kinematic constraints, to only include stars most likely to be in binary pairs based on the measured kinematics and the signal-to-noise (hereafter, SNR) of the measurements: 

\begin{enumerate}
\item The SNR on the distance to the TGAS star was larger than 5.
\item The total proper motion of the TGAS star was larger than 10 mas yr$^{-1}$.
\item The SNR of the total proper motion, of the TGAS star, was larger than 10.
\item $\Delta d < min(1\sigma_{\Delta d},100$~pc$)$, where $d$ is the distance in pc with $\sigma_d$ representing 1 standard deviation in the distance error. 
\item $(\frac{\Delta\mu_{\alpha}}{\sigma_{\mu_{\alpha}}})^2+(\frac{\Delta\mu_{\delta}}{\sigma_{\mu_{\delta}}})^2 < 2$, where $\mu$ is the proper motion; $\sigma$ is the quadrature sum of the errors; and $\Delta\mu$ is the scalar difference between the pair's proper motion components. 
\end{enumerate}

\noindent This led to a final set of 8,781 TGAS-TGAS candidate pairs.

\subsubsection{SDSS Query\label{subsec:sdss}}
We queried the SDSS DR-13 using CASJobs within 3\arcmin of each Tycho-2 position using the STAR, PROPERMOTIONS and TWOMASS tables in CasJobs to create our second data subset (hereafter, TGAS-SDSS). We elected to maintain the strict 3\arcmin of SLW I\& II search radius because of the colossal size of the SDSS dataset ($>400\times10^6$ sources). While our search radius in physical distance will limit the possible detected very wide binary companions for the closest TGAS stars (separation $< 0.5$~pc for a TGAS object at 500~pc), increasing the search radius or using the evolving search radius described in \S~\ref{subsec:tgas}, would return an enormous amount of secondary candidates before vetting ($>10,000$ in some cases). The computational resources necessary for such an analysis is beyond our capabilities and the risk of a substantial number of interloper companions is too high.

We apply the following quality cuts to remove matching SDSS point sources with poor photometry. First, we removed stars from the search based on the following quality flags for \textit{riz}, requiring all to be zero: PEAK$\_$CENTER, NOTCHECKED, PSF$\_$FLUX$\_$INTERP,~INTERP$\_$CENTER, BAD$\_$COUNTS$\_$ERROR, SATURATED, BRIGHT, NOBLEND and DELEND$\_$NOPEAK. We only required these cuts on the \textit{riz} magnitudes because these were the magnitudes we used to identify varying spectral types and distances. Second, we set limits on the size of the residual difference between the \textit{riz} point-spread-functions (hereafter, PSF), requiring $PSFMAG_r - PSFMAG_i >= 0.30$, $PSFMAG_i=PSFMAG_z >= 0.20$. Third, we required each PSF to be larger than 0 with the absolute value of the uncertainty to be less than 0.1~mag. Finally, we required the extinction in \textit{r} to be less the 0.5 magnitudes and for the proper motion be larger than 40 mas yr$^{-1}$. This search resulted in 195,212 candidate secondaries from the SDSS catalog for 136,969 TGAS stars. 

Finally, we included the 5 kinematic TGAS quality cuts described above to further constrain our sample. The addition of these cuts created our final sample of 160 TGAS-SDSS candidate pairs.

\subsubsection{Spectral Type Relations and Distances with SDSS Photometry\label{subsec:photrelations}}

We estimated the spectral types of our stars, and the distances to SDSS candidate objects, following the procedure from previous SLoWPoKES work, by using SDSS photometric colors \citep{Dhital:2010, Dhital:2015}. For stars assumed to be on the main sequence, whose colors did not match a sub-dwarf or white dwarf, we use the relation from \citet{Covey:2007}. For stars with main-sequence colors that match K5 to M9 we use the relation from \citet{Bochanski:2010}. We estimated the distances for stars with colors matching spectral types later than M9 using the relation from \citet{Schmidt:2010}. 

For stars whose colors match those of subdwarfs, we use the relation from \citet{Bochanski:2012}. We calculate the distance to stars with the colors of white dwarfs using \textit{ugriz} photometry and the \citet{Bergeron:1995} models. These models use a chi-square minimization technique and assume a hydrogen dominated atmosphere. Any star without a color matching to the above categories was removed from our sample.

\subsection{Assessing False Positive Likelihood with a Galactic Model\label{sec:select}}

The cuts described in \S~\ref{sec:identify} provide a sample of candidate wide binary pairs but provide little evidence to the authenticity of each pair. Consequently, many of these candidates could be due to chance alignments. While the inclusion of the \textit{Gaia} parallaxes helps to alleviate this tension, the uncertainties in the parameters used to determine the SDSS color based distances do not completely remove ambiguity. This is particularly important for pairs with large angular separation. 

Similar to previous SLoWPoKES work, we employ a 5-dimensional model to recreate the stellar populations along a given line of sight and calculate the probability of a chance alignment. This model applies the empirically measured parameters of the Milky Way to account for stellar number density and spacial velocities to asses the chance of alignment probability. We provide a brief summary of the model below and direct the reader to \citet{Dhital:2010} for a more detailed explanation.

An ideal model Galaxy would be simulated with $10^{11}$ stars based on the number density and population of the Milky Way each with a unique position, proper motion and radial velocity. However, given the enormous computational resources required for such a simulation, our model instead simulates a cone, centered at the ($\alpha$, $\delta$) of each TGAS source out to a distance of 2.5~kpc. The stellar number density in the cone is calculated by integrating the Galactic density profiles and assumes a bimodal disk with an ellipsoidal halo. The line of sight is repopulated with stars using a rejection method to ensure a random redistribution using the previously calculated stellar density. For the binary stars in the GAMBLES sample, typically between 150 and 15,000 stars were generated along a given line of sight for each $\alpha$, $\delta$ and distance. While this two-dimensional model is adequate for our needs we stress it is an oversimplification of the Galactic scale height and does not reproduce local variation such as moving groups or clusters.

We then ran an ellipsoidal search, defined by the angular separation, distance errors and proper motions of a given pair, to simulate our SDSS and TGAS searches described above. 1,000 Monte-Carlo realizations were run for each pair. We divided the total number of stars along a given line of sight, with matching characteristics to our candidate binaries, by the total number of Monte-Carlo realizations to determine how likely a given binary pair was due to a chance alignment. 

We stress that this metric is not a formal probability but instead an estimate on the number density of stars with similar characteristics in a given 5-D Voxel (hereafter, $V_5$). Stars which returned a value of $V_5=0$ (i.e. $<1$ in 1000 stars) where assigned $V_5=0.001$ because their true $V_5$ was below the resolution of our Monte-Carlo procedure. We selected candidate pairs which return a value of $V_5 \le 0.05$ as \textit{bona-fide} wide binary pairs following previous SLoWPoKES work \citep{Dhital:2010, Dhital:2015}.

\begin{figure}[ht]
    \centering
    \includegraphics[width=\linewidth]{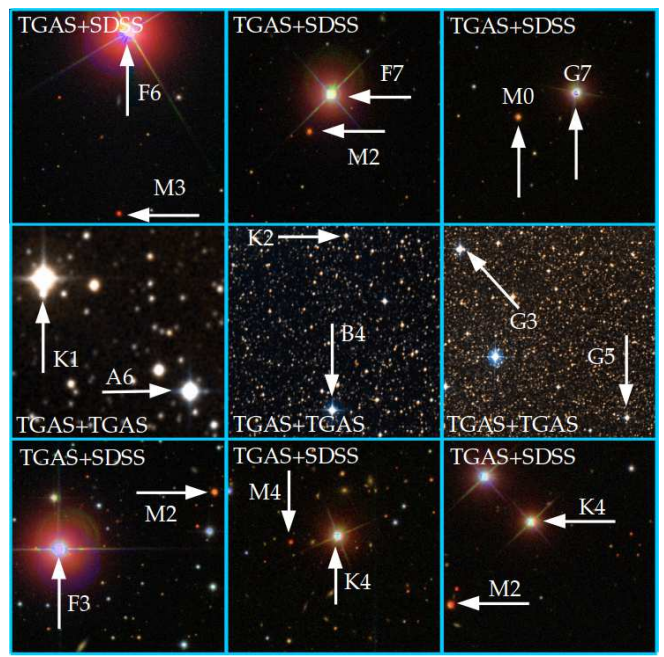}
    \caption{SDSS images (TGAS-SDSS binaries) and DSS images (TGAS-TGAS sample) of 9 selected binaries from the GAMBLES sample. Images from SDSS are 3\arcmin on a side, while images from DSS vary in size from $2-4\arcmin$, reflecting the large angular separations of some of the TGAS-TGAS binary candidates. The label on each frame details whether the pair was discovered by a match between TGAS and SDSS or within TGAS.}
    \label{fig:pics}
\end{figure}

\section{Results\label{sec:results}}

\subsection{A Catalog of TGAS-TGAS and TGAS-SDSS Wide Binaries}

\begin{figure*}[ht]
    \centering
    \includegraphics[width=\linewidth]{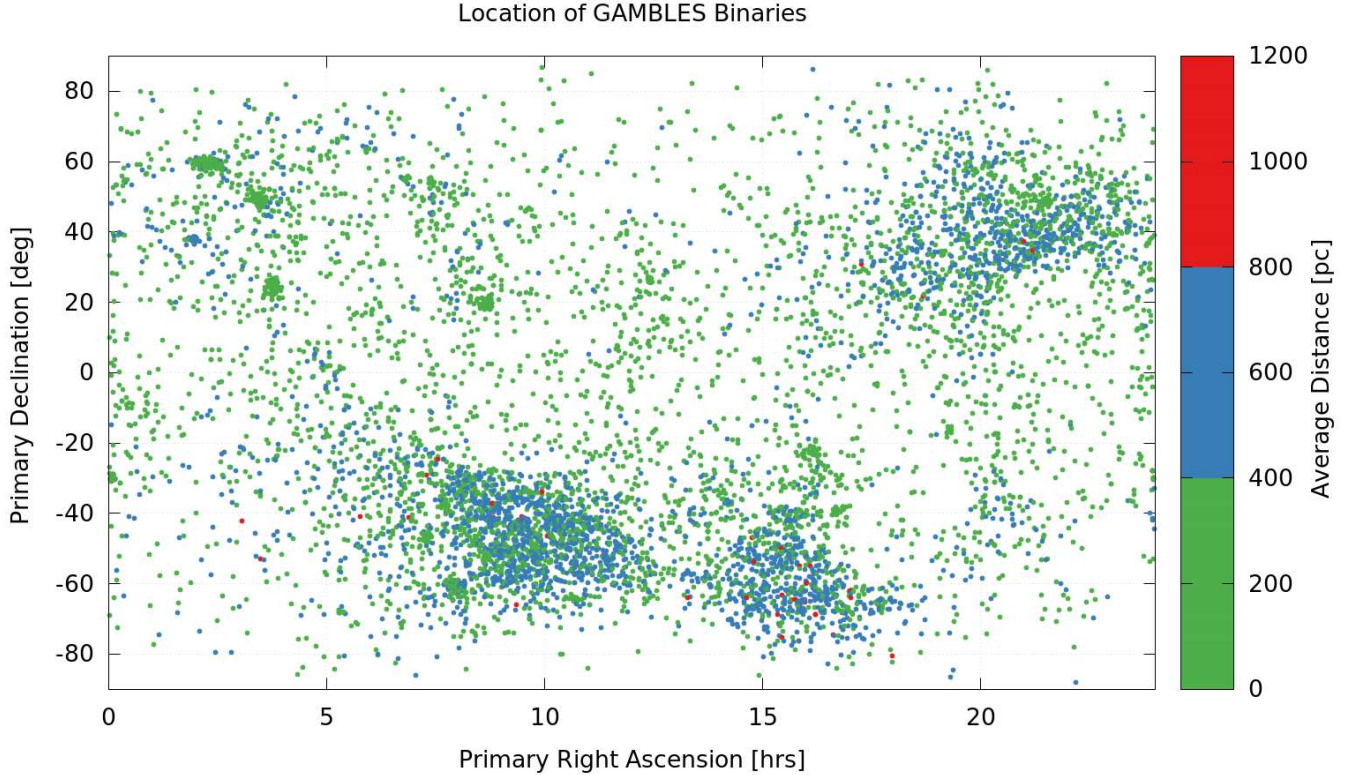}
    \caption{An all-sky map of GAMBLES pairs with statistically similar kinematics as reproduced in our Galactic Model. The map shows position of the primary star on the sky in Right Ascension [hours] and Declination [deg]. The color denotes the distance to the binary from the Sun, in pc.}
    \label{fig:location}
\end{figure*}

\begin{figure*}[ht]
    \centering
    \includegraphics[width=.75\linewidth]{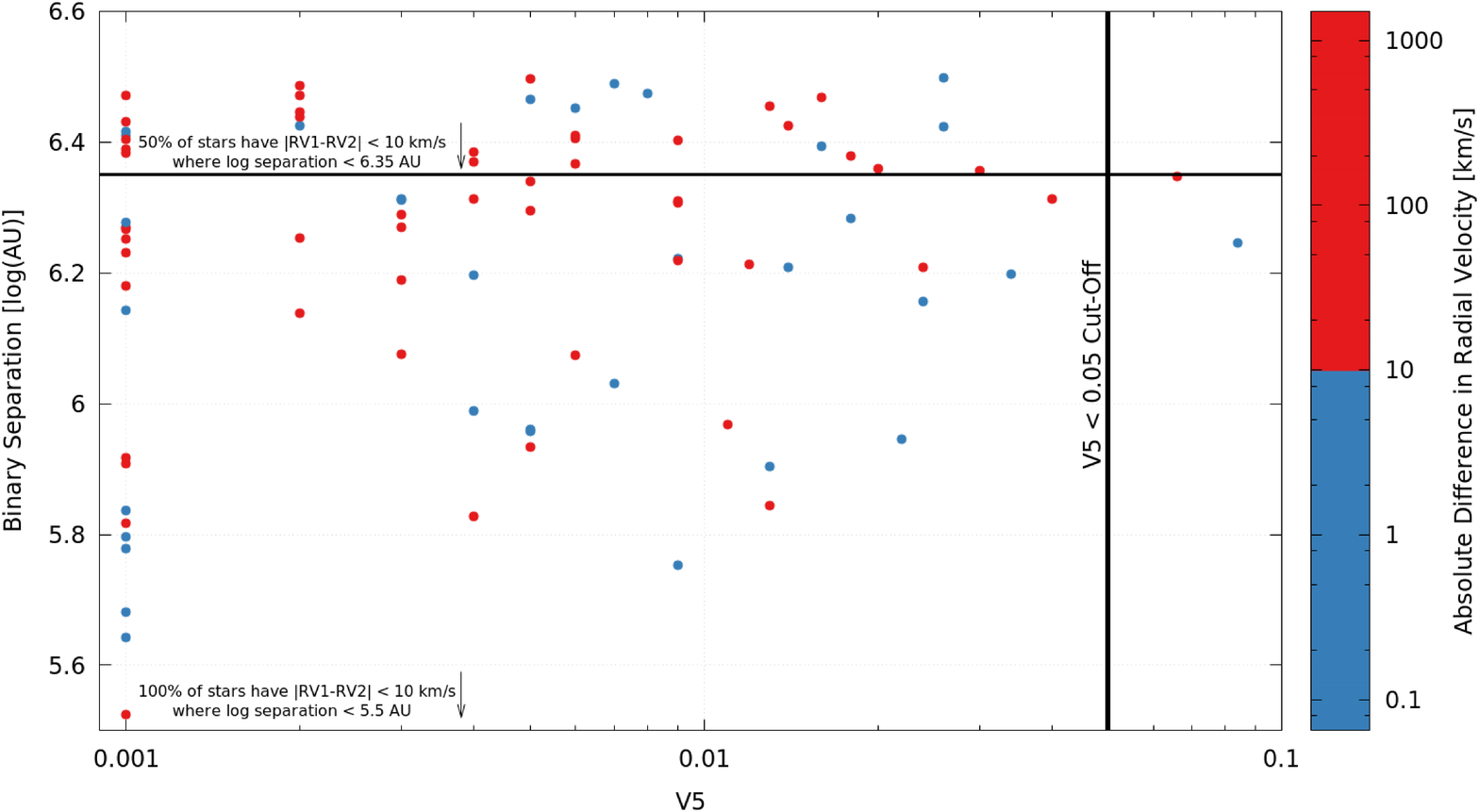}
    \caption{A comparison of the $V_5$ statistic and binary pairs showing consistent radial velocities ($\abs{RV_1-RV_2} \le 10$). We find for pairs with separations closer than $10^{5.5}$~AU, $100\%$ of objects show similar RV values. However, this percentage drops to $50\%$ for stars with separation closer than $10^{6.35}$~AU. This suggests that the $V_5$ statistic is a reasonable metric for determining pairs out to separation of $10^{5.5}$~AU but is less reliable at larger separations.}
    \label{fig:cuts}
\end{figure*}

Our requirement of $V_5\le0.05$ produced 8,660 statistically significant wide-binary pairs with separations of $10^3-10^{6.5}$~AU: 144 out of 160 candidate pairs in the TGAS-SDSS sample and 8,516 out of 8,781 candidate pairs in the TGAS-TGAS sample. Figure ~\ref{fig:pics} shows photographic cut-outs of SDSS \& DSS images for 9 statistically significant wide-binary pairs, while Figure~\ref{fig:location} shows the location of the candidate binaries in celestial coordinate space. 

We tested the reliability of our $V_5$ statistic using measured radial velocity information for binary candidates in our sample. We identified 92 pairs from the TGAS-TGAS sample with reliable radial velocity measurements (RV$_{SNR} >1$) across 8 large-scale spectroscopic surveys using information provided in the TESS Input Catalog \citep{Stassun:2017}: Geneva-Copenhagen, Gaia-ESO, GALAH, RAVE, LAMOST, APOGEE, SPOCS \& PASTEL \citep{Holmberg:2009, Gilmore:2012, Kordopatis:2013, DeSilva:2015, Luo:2015, Majewski:2015, Brewer:2016, Soubiran:2016}. When more than one catalog provided information for both stars we accepted the preferred provenance provided by the TESS Input Catalog.

As shown in Figure~\ref{fig:cuts}, we find all seven candidate pairs with binary separations $<10^{5.5}$~AU have low $V_5$ values ($<0.003$) and consistent RV measurements ($\Delta RV < 10$~km/s). However, the percentage of pairs with similar RV measurements drops past this separation and becomes closer to 50\% near binary separations of $10^{6.3}$~AU. We consider the $V_5$ statistic to be a satisfactory indicator of true wide binary pairs for separations out to $\sim10^{5.5}$~AU but pairs with larger separations will need to be subjected to further scrutiny for validation. Nevertheless we keep all pairs passing our statistical cut for the remainder of our analysis.

We have made the entire GAMBLES catalog, including the candidate binaries that did not pass our the scrutiny of our Galactic Model and the analysis described in \S~\ref{subsec:longlive}, available on the Filtergraph data portal and visualization system\footnote{\url{https://filtergraph.com/gambles}} \citep{Burger:2013}. This allows the catalog to be easily updated as future {\it Gaia} and other catalogs become available, enables users to make their own quality control cuts that may differ from those we have used here, and provides ease of access for any follow up observations to further investigate any specific binary pair.

\subsection{Distribution of Binary Component Types \label{subsec:dist_mass}}

\begin{figure}[ht]
    \centering
    \includegraphics[width=\linewidth]{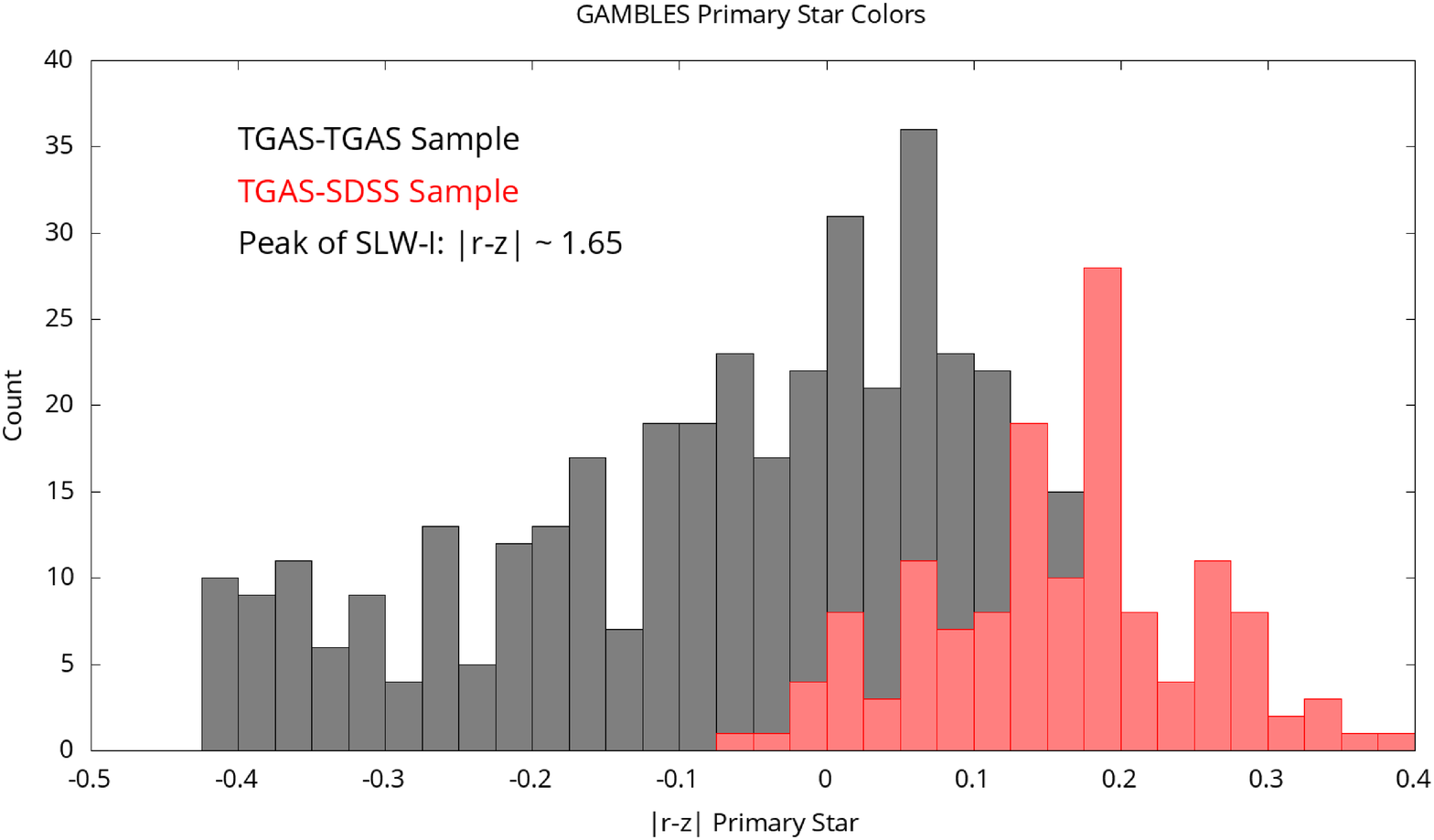}
    \includegraphics[width=\linewidth]{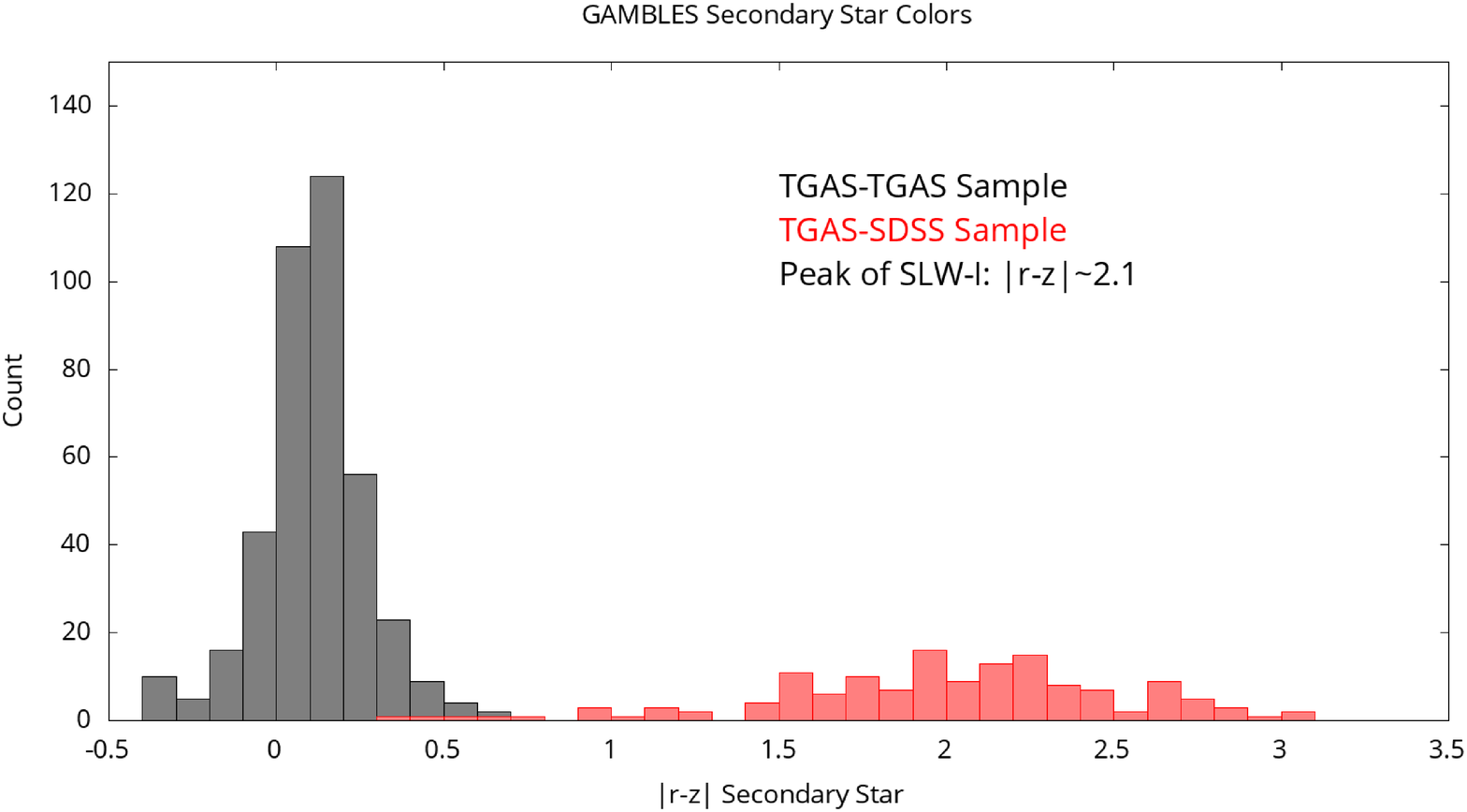}
    \caption{The $\abs{r-z}$ color for the GAMBLES candidate binaries primary stars (\textit{Top}) and secondary stars (\textit{Bottom}). The $\abs{r-z}$ color can be used as a proxy for mass with bluer colors denoting higher mass stars and redder colors denoting lower mass stars. Overall, the GAMBLES sample is higher mass (peak $\abs{r-z}\sim$0.05) than SLW-I \& SLW-II (peak $\abs{r-z}\sim$1.5). The TGAS-SDSS sample is composed of lower mass binary pairs relative to the TGAS-TGAS sample and the secondary stars in the TGAS-SDSS sample are nearly exclusively low mass. For visual clarity, we only display the subset of the GAMBLES sample with dissipation lifetimes $>1.5$~Gyr (see \S~\ref{subsec:final}).}
    \label{fig:bin_mass_pop}
\end{figure}

We categorized the 8,660 assorted mass binary pair candidates into a basic class and spectral type using the color relations described in \S~\ref{sec:identify}. Each star was assigned a spectral type, O to M, and given a basic classification: main-sequence (MS), very-low mass (VLM), white dwarf (WD) or sub-dwarf (sdM). We also assigned a spectral type to each binary component if the stellar colors, $\textit{r}-\textit{z}$, were within the acceptable ranges. We find we recover binary component spectral types from O to M. 

Generally we find the pairs in the TGAS-SDSS sample to be of lower mass ($\langle r_1-z_1\rangle\sim0.15$;$\langle r_2-z_2\rangle\sim1.99$) while the pairs in the TGAS-TGAS sample are of higher mass ($\langle r_1-z_1\rangle\sim-0.05$;$\langle r_2-z_2\rangle\sim0.11$) as is shown in Figure~\ref{fig:bin_mass_pop}. This was expected given the TGAS-SDSS sample was selecting secondary stars nearly exclusively from late type stars.

We stress, however, that this spectral typing has been based on photometry alone and there could be some miss-classifications. This is particularly important in the case of the stars in the TGAS catalog, where the photometry is synthetic rather than observed. These categorizations are not statistically scrutinized by our Galactic Model and we cannot quantify their integrity without proper photometric and spectroscopic follow-up. However, these categorizations passed each of our imposed color cuts and we have kept the designations in the final GAMBLES catalog.

\subsection{Distribution of Binary Pair Separations}

\begin{figure}[ht]
    \centering
    \includegraphics[width=\linewidth]{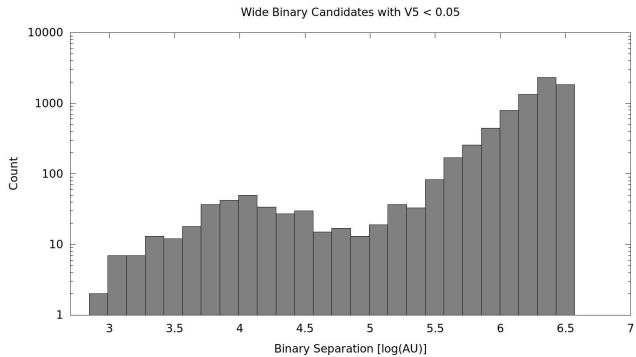}
    \caption{The projected separation for all pairs with $V_5<0.05$ in log(AU). We find the majority of the binaries have separations greater than $10^{5.5}$~AU, where the $V_5$ statistic becomes unreliable (see Figure~\ref{fig:cuts}). These objects likely represent the co-moving pairs mentioned in \citet{Semyeong:2016}.}
    \label{fig:separation}
\end{figure}

The 1,342 SLW-I and 105,537 SLW-II binary pairs were found to have separations between $10^3-10^5$~AU. The GAMBLES sample of 8,660 assorted mass binary candidates greatly extends the SLoWPoKES sample to separations $>10^6$~AU (see Figure~\ref{fig:separation}). While the distributions of binary pairs in our TGAS-SDSS sample is ``tight" with separations of $10^{3.3}-10^{5}$~AU or $0.01-0.54$~pc, we find the TGAS-TGAS sample to be much broader with separations ranging from $10^{2.5}-10^{6.5}$~AU or $0.002-15.5$~pc.

Given the large separation between the TGAS-TGAS components, $>3$~pc in some cases, it would be reasonable to assume the majority of these pairs are not wide binaries \textit{per se} but instead co-moving pairs in larger diffuse stellar associations or moving groups. Recent work by \citet{Semyeong:2016} identified 13,085 common proper motion pairs in the TGAS data set, many with separations $>3$~pc. Since we are only interested in pairs which could be considered gravitationally bound, independent of a larger group or cluster, we have adapted our selection techniques to consider the binding energies, dissipation lifetimes and the reliability of the $V_5$ statistic when selecting our final catalog of wide binary candidates, as we now discuss.

\begin{figure}[ht]
    \centering
    \includegraphics[width=\linewidth]{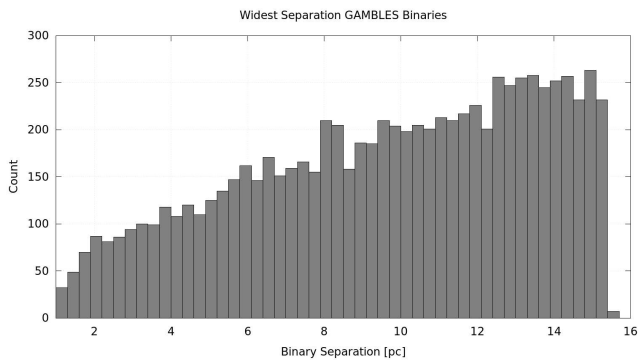}
    \caption{Statistically significant ($V_5<0.05$) GAMBLES binary candidates with very wide separations ($>1$~pc). Many objects with separations $>1$~pc may be co-moving pairs within larger groups or clusters as also identified in \citet{Semyeong:2016}.}
    \label{fig:very_wide}
\end{figure}

\subsection{A Physically Motivated Subset of the GAMBLES Catalog: Long-Lived Binaries\label{subsec:longlive}}

While our Galactic Model does a proficient job of determining chance alignment probabilities in 5 dimensional space, particularly at separations of $<10^{5.5}$~AU, it does not qualify binary likelihood based on the physical properties of each system. It is critical we address the practicality of many of these systems, particularly those with separations $>1$~pc or at heliocentric distances $>2.5$~kpc where our Galactic model does not adequately represent the full volume in a given 5-D voxel along the line-of-sight. 

Given the large separations of the widest binary candidates ($a>1$~pc) in the initial GAMBLES sample (see Figure~\ref{fig:very_wide}), we wanted to objectively test the validity of each pair. This objectivity is particularly important given the statistical nature of our selection because we cannot simply exclude a binary pair if the component separation appears implausible but the pair alignment is statistically significant.

We further investigate the authenticity of each binary pair in two ways. First, we calculate the binding energies of each pair to place a realistic constraint on the furthest separation physically allowed. And second, we determine the expected dissipation lifetimes for each pair to predict and understand their future gravitational stability. 

\subsubsection{Binding Energies of the GAMBLES Sample}

Our Galactic model provides the expected number density of stars along a given 5-D voxel of positions, distances and proper motions. While the model provides sufficient evidence against the chance alignment of similar stars it is not immune to impostor pairs. We calculated the binding energies of the GAMBLES sample to quantify the physicality of many of the widest pairs. We define the binding energy of a pair, measured in ergs, as the gravitational potential energy between the two objects:

$$U = \frac{GM_1M_2}{a}$$

\noindent where G is the gravitational constant, $M_1$ and $M_2$ are the masses of each binary component and $a$ is the distance between the two stars. We convert from angular separation to physical separation following the logic of \citet{Fischer:1992}, which used Monte Carlo simulations probing a variety of binary parameters, to find $a=1.26\Delta\theta d$; where \textit{a} is the physical separation, $\Delta\theta$ is the angular separation and $d$ is the distance to the pair. Figure~\ref{fig:life_be} shows calculated binding energies for each binary where an $r-z$ color was available to estimate a given component's mass.

All stars from the SDSS sample were required to have suitable $r-z$ colors in order to calculate the distance to the star. The distances for the TGAS stars did not require SDSS photometry because each star had an observed \textit{Gaia} parallax. While we used the synthetic photometry of \citep{Pickles:2010} to estimate stellar mass, not all stars in the TGAS-TGAS sample had available synthetic photometry. We have elected to remove the 985 pairs from the final GAMBLES, long-lived sample which could not have their mass, and thus their binding energy, estimated.  

Not surprisingly, many of the candidate GAMBLES binaries, with separations $>10^{5.5}$~AU, appear to break the previously observed empirical limits for stellar binding energy: $10^{40}$ for sub-stellar objects; $10^{41}$~ergs for stars; and $10^{42.5}$ for the lowest mass objects \citep{Reid:2001, Close:2003, Burgasser:2003, Close:2007}.

If we impose a hard limit of $10^{41}$~ergs we find that 7,263 of the binaries with calculable masses fall below this limit. The fact that these binaries have binding energies below what is expected, but are statistically significant in our Galactic Model, has a few implications. The first is that the population of binaries which fail the cut are dynamically unstable and are drifting apart but retain some characteristics of the initial system (such as similar proper motions). This argument is strongly supported by the work of \citet{Semyeong:2016}, which identified $>10,000$ candidate co-moving pairs in the TGAS data set.

The second implication could be that the previous empirical limits were too conservative for the widest binary pairs. The SLW I \& II samples found nearly the entire catalog violated these empirically imposed limits, with binding energies closer to that of Neptune and the Sun at $10^{40}$~ergs \citep{Dhital:2010, Dhital:2015}. 

The third implication is that these candidate binaries have low binding energies because they are not \textit{bona-fide} wide binaries. Their large angular separations, $>2^{\circ}$ in some cases, vastly increases the number of stars any given TGAS-TGAS pair could be matched to. We calculate and interpret the dissipation lifetimes for the remaining 7,736 stars (with $V_5<0.05$ and calculable binding energies) to alleviate the tension created by such wide angular separations.

\subsubsection{The Dissipation Lifetimes of the GAMBLES Sample \label{subsubsec:lifetime}}

\begin{figure*}[ht]
    \centering
    \includegraphics[width=.9\linewidth]{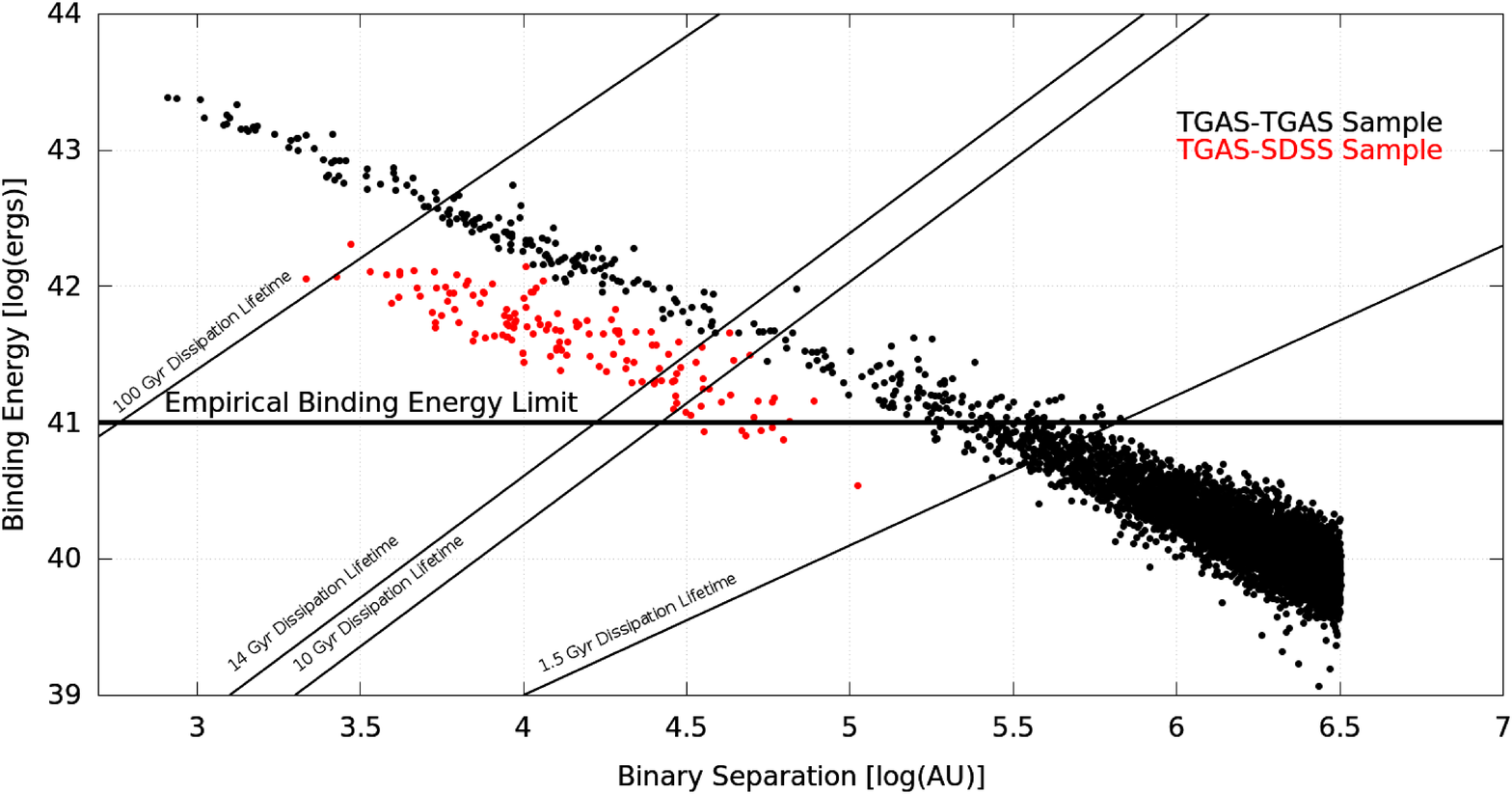}
    \includegraphics[width=.9\linewidth]{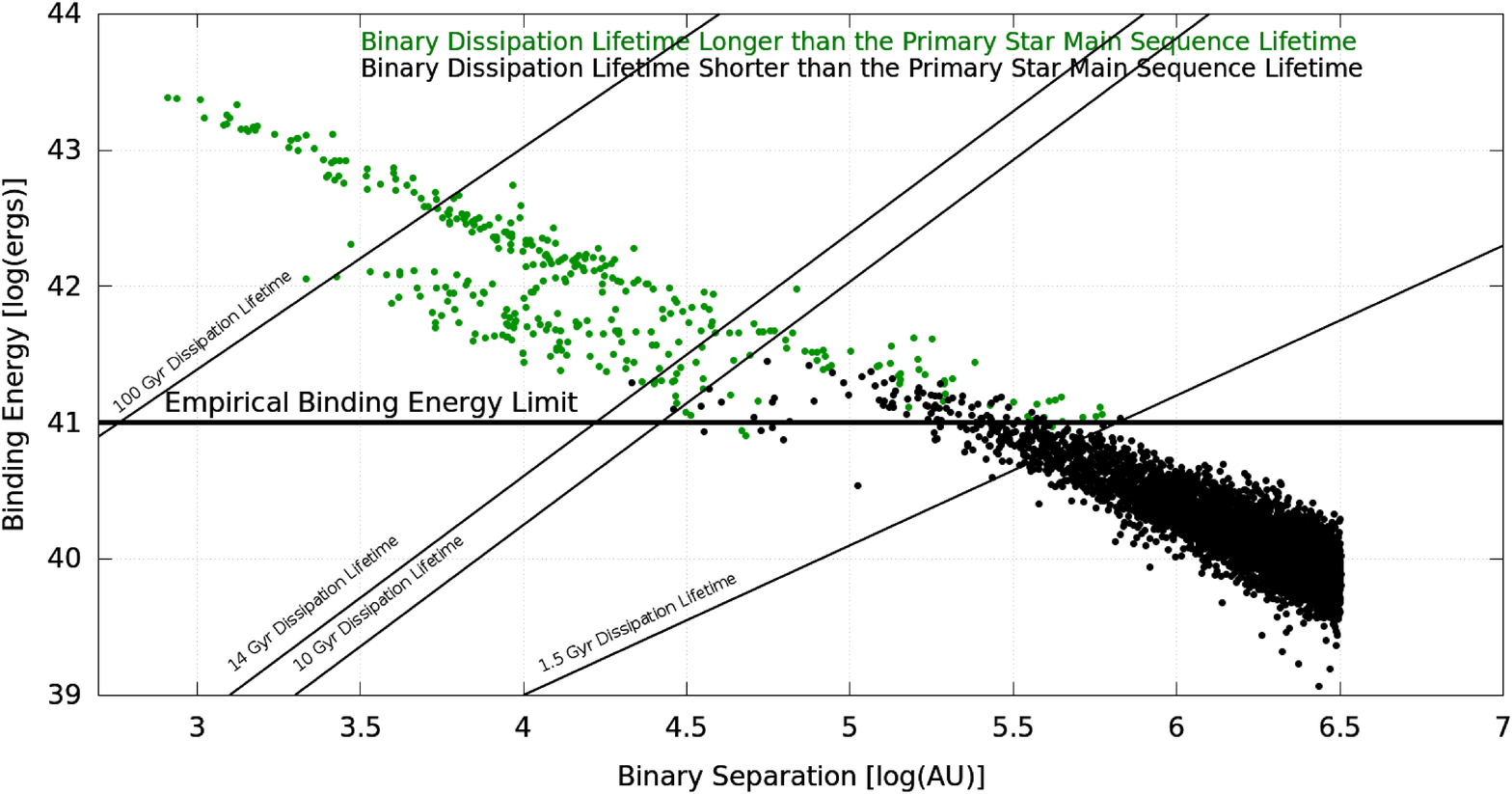}
    \caption{The full distribution of high significance pairs ($V_5< 0.05$) in the GAMBLES sample where a binding energy could be calculated. The over-layed lines denote the estimated dissipation lifetime cut-offs. \textit{Top}: The binding energies for each GAMBLES pair in the TGAS-TGAS sample (black points) and the TGAS-SDSS sample (red points). The gap between the two samples is largely due to the TGAS-TGAS sample generally consisting of higher mass components than the TGAS-SDSS sample. \textit{Bottom}: The same figure as above but with the pairs where the expected dissipation lifetime is longer than the expected main-sequence lifetime of the higher mass component (green dots), suggesting the pairs are bound over long timescales; based on these objects we adopt a cut off of $>1.5$~Gyr for our long-lived wide binary subset.}
    \label{fig:life_be}
\end{figure*}

Even if a binary pair possesses the requisite binding energy to remain stable, the large separation coupled with the local Galactic environment could cause the pair to dissipate over time. Encounters with other stars, molecular clouds or even subtle changes in the overall Galactic potential can combine to disrupt the system's stability and cause it to break apart \citep{Weinberg:1987}. We calculated the average dissipation lifetime of a given binary to aid in our interpretation of the possible stability of each system. We followed the logic of \citet{Weinberg:1987, Close:2007, Dhital:2010} and used the below approximation, based on the advection and diffusion of orbital binding energy due to small encounters, to calculate the expected dissipation lifetimes:

$$a \approx 1.212\frac{M_{tot}}{t_{*}}$$

\noindent where $a$ is the separation in pc, $M_{tot}$ is the total mass of the system in solar masses and $t_*$ is the lifetime of the binary in Gyr. Figure~\ref{fig:life_be} shows the binding energies and separations of each statistically significant GAMBLES pair with the over-plotted dissipation lifetimes of 1.5, 10, 14 and 100~Gyr.

We find the majority of GAMBLES binary candidates have dissipation lifetimes shorter than 1.5~Gyr and binding energies less than $10^{41}$~ergs. In fact, when we compare the dissipation lifetimes of each candidate to the expected estimated main-sequence lifetime of the higher mass component (hereafter, PMSL), we find any star with a dissipation lifetime longer than the PMSL also has a dissipation lifetime longer than 1.5~Gyr and almost exclusively has a binding energy larger than $10^{41}$~ergs.

However, there is overlap between stars with dissipation lifetimes shorter than the PMSL but larger than 1.5~Gyr and stars with binding energies $<10^{41}$~ergs but dissipation lifetimes longer than 1.5~Gyr. We therefore make the GAMBLES subset of long-lived binaries using a cut off of 1.5~Gyr in dissipation lifetime. This cut retains only the pairs expected to remain bound over long timescales and avoids removing any pair which could be bound for the main-sequence lifetime of the primary.

\subsubsection{Final Sample of Physical Long-Lived Wide Binaries \label{subsec:final}}

These cuts result in 543 binary candidates, which we believe, are the most likely to be genuine wide-binary pairs in the GAMBLES sample that are also likely to survive for timescales of $\gtrsim$1.5~Gyr. We summed the expected false positive values (V$_5$) for these 543 binaries and found the expected number of false positive binaries to be $\Sigma V_5 = 2.4$. Thus this high-fidelity binary population appears to be very robust, even at very wide separations of 1~pc or greater. 

We also searched the GAMBLES long-lived sample for possible systems with more than 2 components. We did this by identifying multiple pairs with an identical primary star. We found 22 systems where the primary was linked to 2 secondary objects and 1 system where the primary was linked to 3 secondary objects. Each of the 23 pairs were found in the TGAS-TGAS sample.

Table~\ref{tb:cand_obs} provides the position and magnitude information for each long-lived wide binary in the GAMBLES sample; Table~\ref{tb:cand_ast} provides the proper motion, distance and Galactic model information for each long-lived wide binary in the GAMBLES sample; and Table~\ref{tb:cand_cal} provides the estimated physical quantities for each long-lived binary in the GAMBLES sample. Quantities for GAMBLES binaries not in the final sample, can be found on the \textit{Filtergraph} portal with the URL \url{https://filtergraph.com/gambles}.

\section{Discussion \label{sec:discussion}}

\subsection{Possible Sources of Uncertainty in GAMBLES}

\subsubsection{The Effect of Giant Contamination on Mass Estimations}

\begin{figure}[ht]
    \centering
    \includegraphics[width=.9\linewidth]{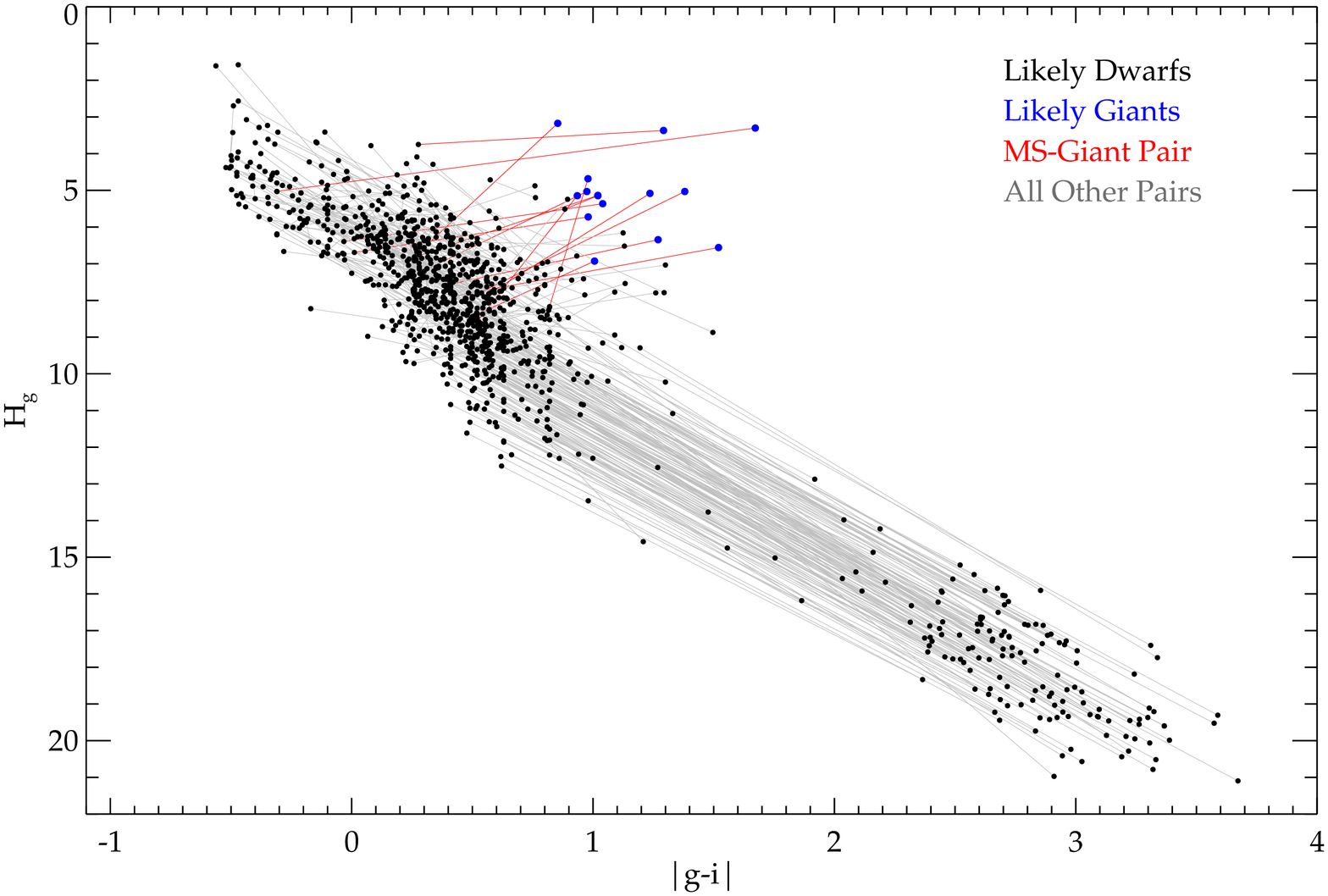}
    \includegraphics[width=.9\linewidth]{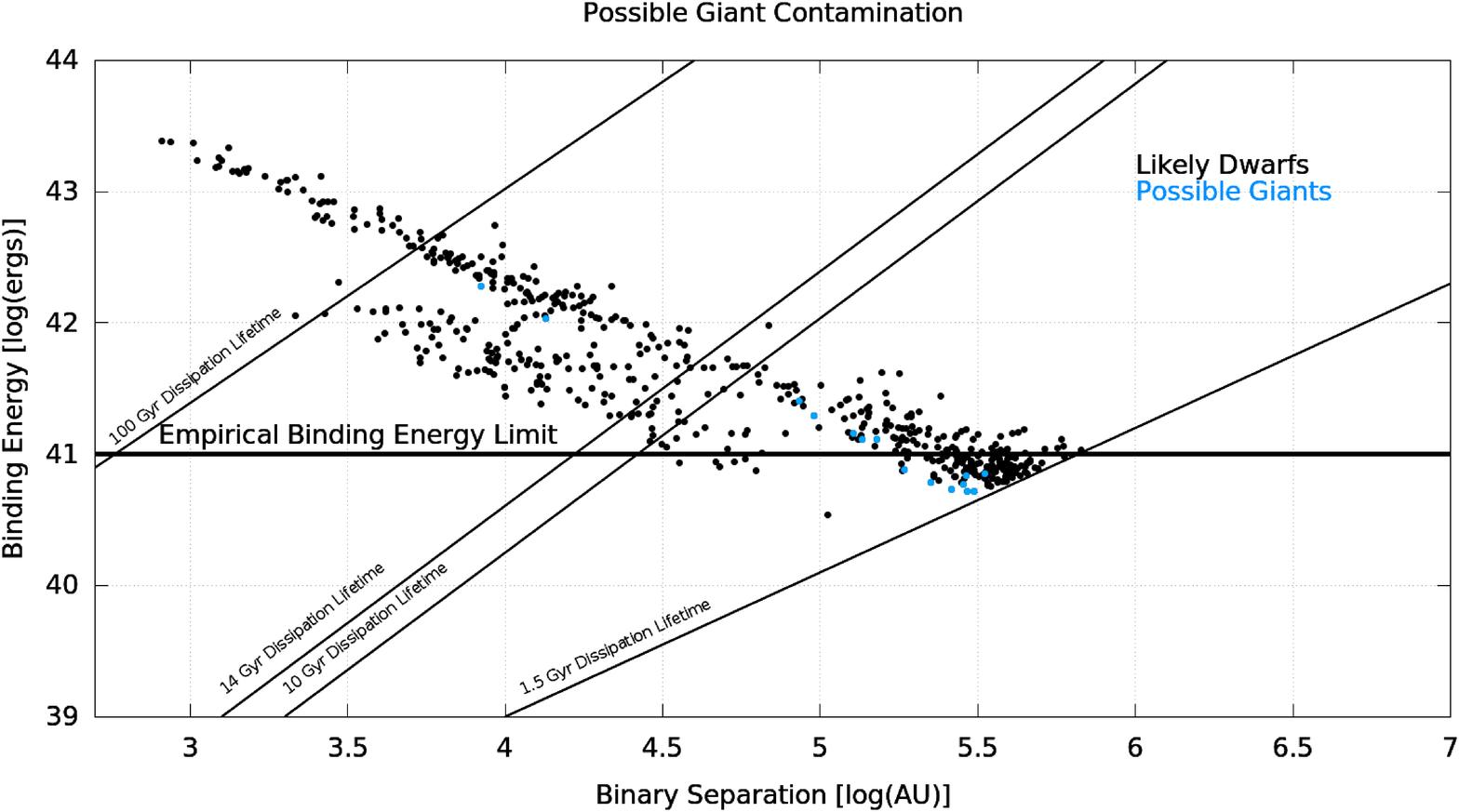}
    \caption{(\textit{top}:) The reduced proper motion diagrams for GAMBLES binaries in \textit{g} used to identify possible giants. Using the relations from \citet{Casagrande:2010} and the \textit{Gaia} distances, the radius of the TGAS stars were estimated. Any star with a radius $>10R_{\odot}$ was flagged as a possible giant (blue dots; likely dwarfs are denoted by black dots). The lines denote matching pairs with grey lines matching between two dwarf stars and red lines matching between a dwarf and likely giant. (\textit{Bottom}:) The binding energies of GAMBLES binaries with at least 1 suspected giant component (blue dots). These binding energies should be treated as upper limits.}
    \label{fig:rpm}
\end{figure}

Figure~\ref{fig:rpm} shows the reduced proper motion in \textit{g} ($H_g$) of each GAMBLES long-lived pair and their respective $g-i$ colors. These reduced proper motion diagrams can provide useful insight into the nature of common proper motion binaries and have been used previously to identify candidate pairs \citep{Chaname:2004, Dhital:2010}. The top part of Figure~\ref{fig:rpm} shows reduced proper motion diagrams for the GAMBLES sample and the link between each binary pair. 

We used the bolometric flux and effective temperature relations from \citet{Casagrande:2010}, coupled with the $B_T$ and $V_T$ magnitudes, to estimate each stellar radii for stars in the TGAS-TGAS, long-lived sample. If a star was shown to have a radius $R > 10R_{\odot}$ we flagged the star as a likely giant. We identified 15 candidate binary pairs where at least 1 star had an estimated radius larger than $10R_{\odot}$. These binary pairs may have decreased binding energies (see bottom panel of Figure~\ref{fig:rpm}) and these values should be considered upper limits. However, if these giant components are in \textit{bona-fide} wide-binary pairs, they would represent the first known systems with evolved components. Because of the small number of possible giants in our sample ($\sim2.8\%$) we do not believe giant contamination largely affects the mass estimates in the GAMBLES population.

\subsubsection{The Effect of Short Period Binaries on Mass and Distance Calculations}
It is possible some of the GAMBLES components are also in short separation systems, unresolved in the SDSS or \textit{Gaia} pixel scales. These short period companions can affect our analysis in a few ways.

First, the resolved, short period companions could lead to an over estimation of the masses of the GAMBLES components by increasing the total flux in a given aperture. This means the dissipation lifetime and binding energy of the system would be overestimated and could artificially force some pairs near the edge of our cuts in to our sample. Second, these short period components could cause the low-mass secondaries from the TGAS-SDSS sample to have incorrect $r-z$ colors leading to incorrect distance estimates. Finally, if the binary isn't fully resolved but creates a slight change in the centroid of a given GAMBLES star, the astrometry for that star could be incorrectly calculated resulting in an improper parallax or proper motion measurement. We believe these short period companions would mainly affect the astrometry of the lowest SNR objects, which we have already removed from our sample as discussed in \S~\ref{subsec:tgas}. However, improved astrometric measurements in the second \textit{Gaia} data release will help to alleviate these concerns.

\subsubsection{The Effect of the Malmquist Bias on the Detection of Low-Mass Companions}
The TGAS data set is composed of $\sim2\times10^6$ \textit{Tycho-2} stars and $\sim9\times10^5$ \textit{Hipparcos} with a limiting magnitude near $V<11.5$ and represent some of the brightest stars in the sky. This sample is also inherently close to the Sun, with the maximum distance of any star in TGAS with a parallax SNR$>5$ (the SNR requirement we adopt in this work) being $952\pm 158$~pc  \citep{Hog:2000, Lindegren:2016}. This could lead to inappropriate interpretations of underlying wide binary population statistics in the following way.

It is possible our search for secondary stars, within the TGAS sample, does not recover objects at similar distances to the TGAS primaries but have apparent magnitudes fainter than the faint limit of \textit{Gaia} but brighter than the saturation limit of SDSS. These objects would not have their parallax measured for the first \textit{Gaia} data release. The addition of these objects may shift the secondary peak in the TGAS distributions to the left, as we find more objects with lower mass and tighter separations. We found including object with lower SNR tended to dilute the bimodal signal but because we could not effectively characterize these systems, due to their low SNR, we removed them from our sample. We believe we can best characterize this effect after the second \textit{Gaia} data release provides astrometric information for all objects, likely to be in single star systems, down to a magnitude of $V\sim16$.

\subsubsection{The Effect of the 3\arcmin Search Radius on Discovering Very-Wide Binary Candidates in the TGAS-SDSS Sample}

The adaptive search radius used to create the TGAS-TGAS sample, allowed for wide binaries to be detected with physical separations as large as $\sim15$~pc, regardless of the heliocentric distance of the pair. However, as mentioned in \S~\ref{subsec:sdss}, we elected to maintain a strict 3\arcmin~angular separation when searching for binary pair candidates in the TGAS-SDSS sample. While limiting the TGAS-SDSS angular search radius helped to alleviate the enormous number of potential spurious wide binary candidates we identify, it also greatly limited the number of potential \textit{bona-fide} wide binary candidates with large physical separations ($>0.5$~pc) at close heliocentric distances. It is particularly likely this 3\arcmin~limit omits a large number of statistically significant binary candidates in the TGAS-SDSS sample because we find 6,838 statistically significant ($V_5<0.05$) binary pairs with wide physical separations ($a>0.5$~pc) in the TGAS-TGAS sample.

The addition of these ``missing" binaries could potentially shift the peak of the distribution of TGAS-SDSS candidates to wider binary separations and begin to fill the bimodal distribution shown in Figure~\ref{fig:bin_tot_samp}. However, the TGAS-SDSS pairs with large separations will have lower masses than their TGAS-TGAS counterparts (see \S~\ref{subsec:dist_mass}). This will cause many of these potential pairs to have smaller binding energies ($<10^{40}$~ergs) and shorter dissipation lifetimes ($<1.5$~Gyr), which would have eliminated them from our long-lived sample (see \S~\ref{subsec:longlive}).

\subsection{The Diversity of the GAMBLES Sample}

\begin{figure*}[ht]
    \centering
    \includegraphics[width=.9\linewidth]{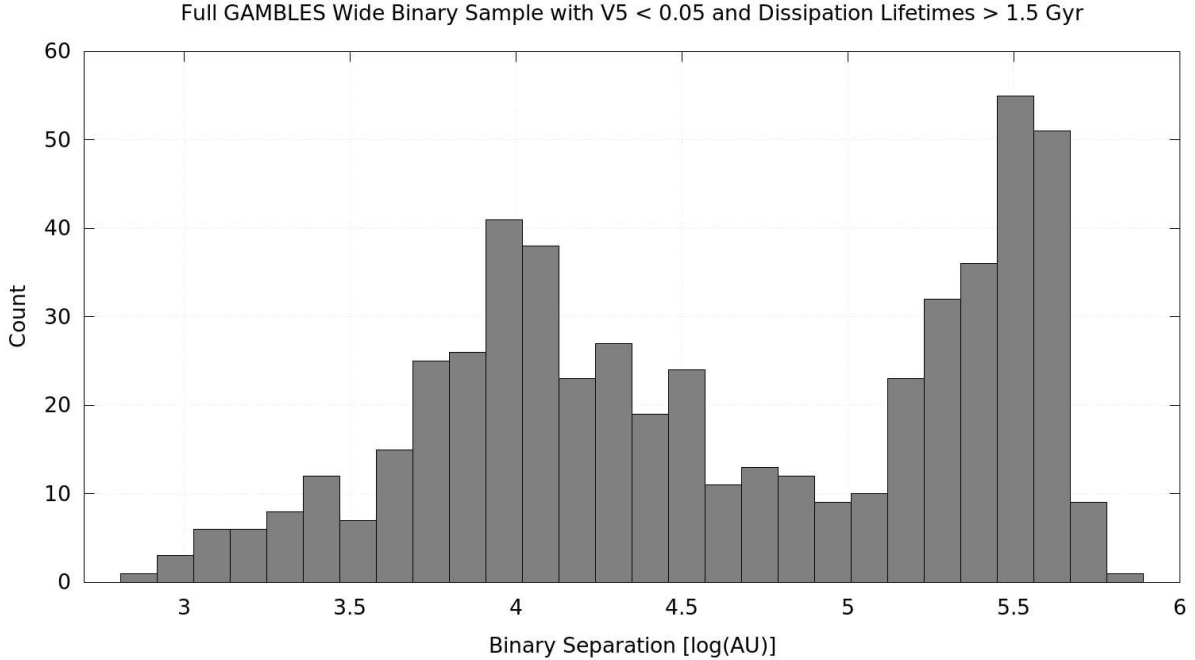}
    \caption{The full distribution of GAMBLES long-lived wide binaries (dissipation lifetimes $>1.5$~Gyr).The distribution is clearly bimodal and is split near binary separations of $10^{4.7}$~AU, suggesting two separate binary populations.}
    \label{fig:bin_tot_samp}
\end{figure*}

\begin{figure*}[ht]
    \centering
    \includegraphics[width=.9\linewidth]{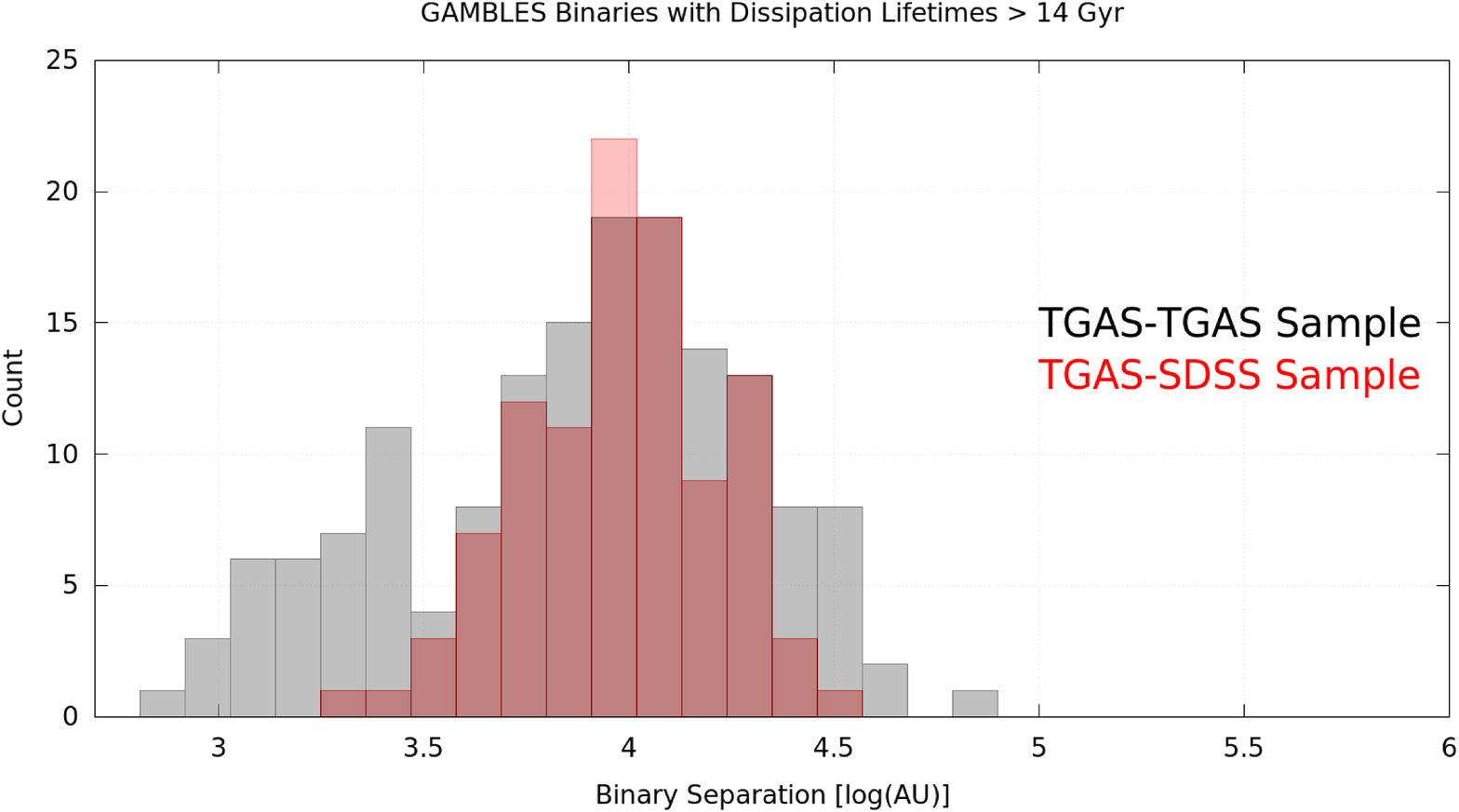}
    \includegraphics[width=.9\linewidth]{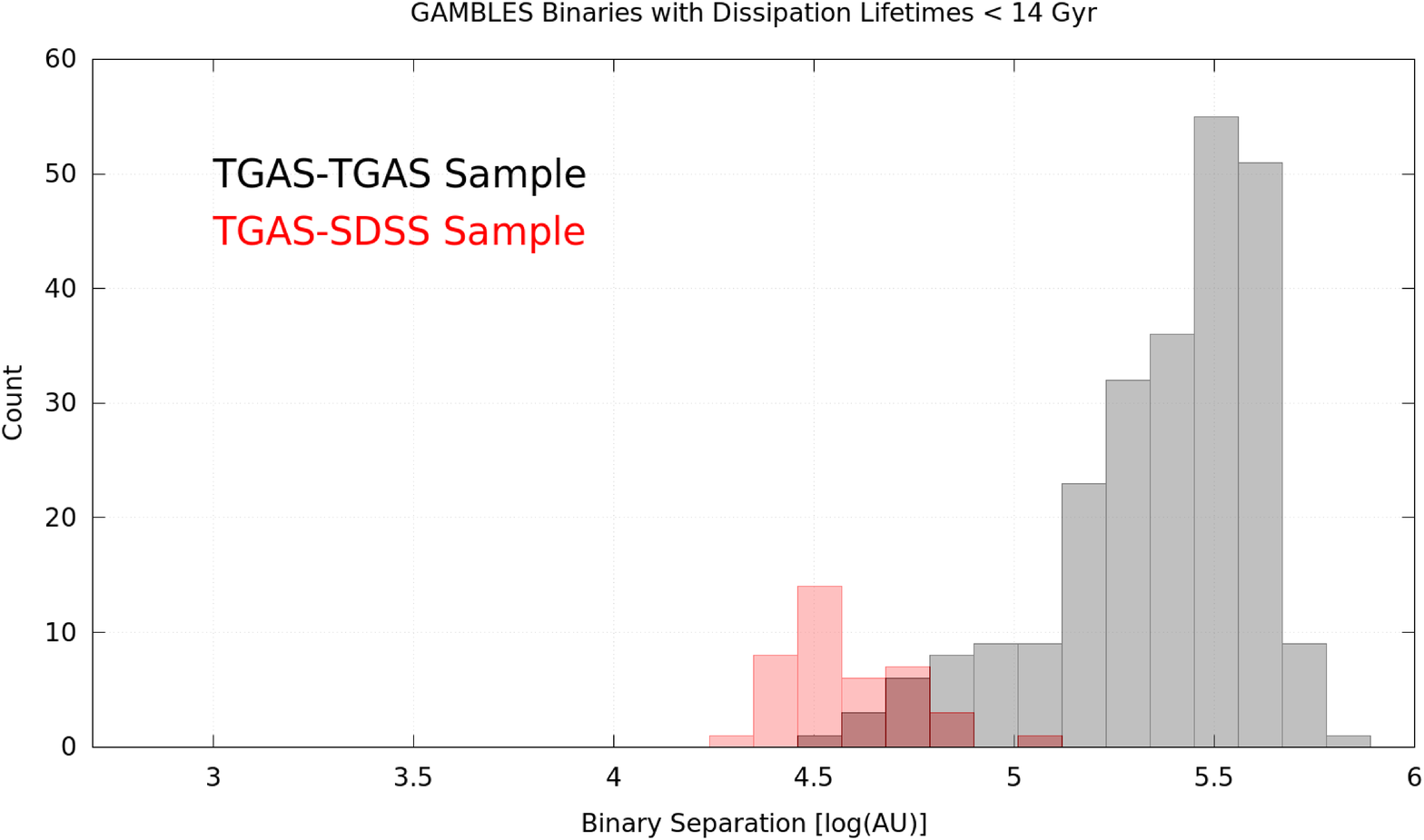}
    \caption{The separation of GAMBLES binaries partitioned by whether the dissipation lifetime of the binary is longer (\textit{top}) or shorter (\textit{bottom}) than 14~Gyr, roughly the age of the universe, separating out the bimodality from Figure~\ref{fig:bin_tot_samp}. The top population is likely binaries which formed in the same molecular cloud and remain bound on long timescales ($>14$~Gyr). The bottom population are likely loosely bound binary pairs that became gravitationally bound within their original cluster environment but will dissipate on shorter timescales. That the peak separation in the bottom panel is different for the TGAS-TGAS versus TGAS-SDSS samples is likely a reflection of the much lower secondary masses probed by the SDSS photometry relative to TGAS; the lower-mass secondaries from the SDSS sample therefore have smaller physical separations for a given binding energy.}
    \label{fig:bin_sep_pop}
\end{figure*} 
As previously stated, the SLW-I catalog identified a bimodal distribution at $10^{3.6}$ and $10^{4.7}$~AU. This was interpreted as a combination of separate wide binary populations in the Galaxy and evidence of varying formation and evolutionary methods for differing binary populations \citep{Dhital:2010}. While the SLW-II catalog did not recover a bimodal distribution in its binary population, it was best described with the combination of two power law functions, likely due to varying combinations of the two SLW-I populations \citep{Dhital:2015}.

The bimodal peak at $10^{3.6}$~AU was suggested to represent a population of wide binaries (hereafter, WBPop-I) which formed together by the same fragmentation of a molecular cloud core but migrated outwards due to dynamic instability, perhaps involving a hierarchical triple-star system structure. This formation mechanism has been shown in recent work by \citet{Reipurth:2012, Elliott:2016}, and indeed a large fraction of the SLW-I wide binaries were later confirmed to be hierarchical triples, with the highest occurrence of triples among the widest of the SLW-I pairs \citep{Law:2010}. Conversely, the bimodal peak at $10^{4.7}$~AU was proposed to represent a population of binary pairs which are loosely bound after the dissipation of the main star forming cluster (hereafter, WBPop-II). This hypothesis was reinforced by the size of a typical pre-stellar core matching the separations of many of the candidate binary pairs ($\sim0.35$~pc or $\sim10^{4.9}$~AU \citep{Kouwenhoven:2010}). 

The GAMBLES catalog shows both similarities and differences to the previous SLoWPoKES catalogs. The GAMBLES long-lived subset clear shows distinct bimodal features, as shown in Figure~\ref{fig:bin_tot_samp}, reminiscent of SLW-I. Additionally, the bimodal distribution of the TGAS-SDSS sample is strikingly similar to the SLW-I sample with peaks in the distribution at $10^4$ and $10^{4.55}$~AU (see Figure~\ref{fig:bin_sep_pop}). 

We further separate the samples to investigate their expected gravitational boundedness with time. We find if we split the GAMBLES sample by dissipation lifetimes longer than or shorter than the age of the universe ($\sim14$~Gyr \citep{Riess:2016}, see Figure~\ref{fig:bin_sep_pop}) the two peaks of the bimodal distribution separate entirely. This suggests WBPop-I objects will stay bound indefinitely. Conversely, WBPop-II objects will eventually dissipate into co-moving pairs (as seen in \citep{Semyeong:2016}) and further dissipate completely into unbound objects.

The WBPop-I peaks (near $10^4$~AU) in the TGAS-TGAS sample and the TGAS-SDSS sample appear to overlap with one another (see the top part of Figure~\ref{fig:bin_sep_pop}). We believe this overlap suggests the formation WBPop-I objects is dependent on the mass of the primary star. Both the TGAS-TGAS and TGAS-SDSS primary stars were drawn from the same TGAS sample and have similar masses (see the top part of Figure~\ref{fig:bin_mass_pop}). The shift of this peak relative to the SLW-I peak ($10^{3.6}$~AU vs. $10^4$~AU) is likely the result of the average GAMBLES primary star mass being larger than the average primary mass in SLW-I. The peak of the $r-z$ color in the SLW-I sample was 1.65, relative to the mean value in the TGAS-TGAS sample of -0.05 and 0.15 in the TGAS-SDSS sample. 

However, the location of the WBPop-II peak in the TGAS-TGAS sample (near $10^{5.5}$~AU) is clearly distinct from the WBPop-II peak of the TGAS-SDSS sample (near $10^{4.5}$~AU) as seen in the bottom part of Figure~\ref{fig:bin_sep_pop}. We believe this suggests the WBPop-II pairs with higher mass secondary components are more likely to stay bound over large distances. We also believe the peak of the population separation for WDPop-II stars may be dependent on the secondary star mass. This is supported by the main difference between the TGAS-SDSS and TGAS-TGAS samples was the selection of secondary components. The TGAS-SDSS sample generally has much lower mass secondary components: the mean $r-z$ color is 1.99 for TGAS-SDSS pairs and 0.11 for TGAS-TGAS pairs. Likewise, the fact that the secondary component masses in the TGAS-SDSS and SLW-I samples are similar, could also explain why the WDPop-II peak of the TGAS-SDSS sample ($10^{4.55}$~AU) is similar to the WDPop-II peak of the SLW-I sample ($10^{4.7}$~AU).

\section{Conclusions\label{sec:conclusions}}

We have extended the SLoWPoKES catalog to include high to medium mass wide binary candidates using the Sloan Digital Sky Survey photometry and parallaxes and proper motions from the \textit{Tycho-Gaia} Astrometric Solution. Our analysis has discovered 8,660 statistically significant, assorted mass pairs, with component spectral types ranging from O to M and separations between $10^3-10^{6.5}$~AU. This data set provides a unique opportunity to study a population of binary stars that is diverse in spectral type and separation.

Our analysis included the comparison of position, distance, angular separation, proper motions, binding energies and dissipation lifetimes to cultivate a subset of 543 GAMBLES long-lived wide binary pairs. We employed a Galactic Model, combined with a Monte Carlo analysis, to determine the expected number of similar stars along a given line of sight. We removed any binary pair from our sample where the number of similar stars per line of sight was $>5$ in 1000. We also excluded any binary star where a mass could not be estimated or the dissipation lifetime was $> 1.5$~Gyr.

We find the GAMBLES sample is split into two populations of wide separation binaries. The first population, with separations near $10^4$~AU, likely formed through outward migration due to dynamic instabilities. Their separations are likely linked to the mass of the primary component and the pair is expected to be bound for their entire stellar lifetime because their dissipation timescales are longer than 14~Gyr. The second population, with separations between $10^{4.5}-10^{5.5}$~AU, likely formed through weak gravitational interactions after cluster dissipation. Their separations could be linked to the mass of the secondary component and these pairs will likely dissipate into co-moving pairs and finally unbound objects on shorter timescales $t_{*}<14$~Gyr. 

Our results corroborate previous SLoWPoKES work which found a bimodal distribution of wide binaries and hint at a possible relationship between component mass, binary separation and dissipation lifetime driving the formation and evolution of these systems. The GAMBLES catalog is available on the Filtergraph portal at \url{https://filtergraph.com/gambles}. This web-based service allows for open source access to many datasets and provides fast plotting and filtering features useful in determining data trends and visualizing large data sets \citep{Burger:2013}.

\acknowledgements

The authors would like to thank the anonymous referee for their comments and suggestions which greatly helped to improve this manuscript. RJO would also like to thank Wilfried Knapp for his thoughtful suggestions and constructive criticism. This work has made use of data from the European Space Agency (ESA) mission \textit{Gaia} (\url{http://www.cosmos.esa.int/gaia}), processed by the \textit{Gaia} Data Processing and Analysis Consortium (DPAC, \url{http://www.cosmos.esa.int/web/gaia/dpac/consortium}). Funding for the DPAC has been provided by national institutions, in particular the institutions participating in the \textit{Gaia} Multilateral Agreement. This work has made extensive use of the Filtergraph data visualization service \citep{Burger:2013} at \url{http://www.filtergraph.vanderbilt.edu}. This research has made use of the VizieR catalogue access tool, CDS, Strasbourg, France. This work was conducted in part using the resources of the Advanced Computing Center for Research and Education at Vanderbilt University, Nashville. Funding for the Sloan Digital Sky Survey IV has been provided by
the Alfred P. Sloan Foundation, the U.S. Department of Energy Office of
Science, and the Participating Institutions. SDSS-IV acknowledges
support and resources from the Center for High-Performance Computing at
the University of Utah. The SDSS web site is www.sdss.org. SDSS-IV is managed by the Astrophysical Research Consortium for the  Participating Institutions of the SDSS Collaboration including the 
Brazilian Participation Group, the Carnegie Institution for Science, 
Carnegie Mellon University, the Chilean Participation Group, the French Participation Group, Harvard-Smithsonian Center for Astrophysics, 
Instituto de Astrof\'isica de Canarias, The Johns Hopkins University, 
Kavli Institute for the Physics and Mathematics of the Universe (IPMU) / 
University of Tokyo, Lawrence Berkeley National Laboratory, 
Leibniz Institut f\"ur Astrophysik Potsdam (AIP),  
Max-Planck-Institut f\"ur Astronomie (MPIA Heidelberg), 
Max-Planck-Institut f\"ur Astrophysik (MPA Garching), 
Max-Planck-Institut f\"ur Extraterrestrische Physik (MPE), 
National Astronomical Observatories of China, New Mexico State University, 
New York University, University of Notre Dame, 
Observat\'ario Nacional / MCTI, The Ohio State University, 
Pennsylvania State University, Shanghai Astronomical Observatory, 
United Kingdom Participation Group,
Universidad Nacional Aut\'onoma de M\'exico, University of Arizona, 
University of Colorado Boulder, University of Oxford, University of Portsmouth, 
University of Utah, University of Virginia, University of Washington, University of Wisconsin, 
Vanderbilt University, and Yale University. We acknowledge use of the ADS bibliographic service.

\bibliographystyle{apj}
\bibliography{references}

\clearpage 
\begin{turnpage}

\begin{deluxetable}{ccccccccccccccccc}
\tablefontsize{\footnotesize}
\tablecaption{Positions and Magnitudes of GAMBLES Long-Lived Wide Binaries \label{tb:cand_obs}}
\tablehead{\colhead{GAMBLES ID} &  \multicolumn{2}{c}{Catalog ID} & \multicolumn{4}{c}{Coordinates (J2015)} & \multicolumn{10}{c}{Sloan Magnitudes}  \\
 & \colhead{HIP/Tycho2} & \colhead{SDSS/HIP/Tycho2} &\colhead{$\alpha_1$[hms]} & \colhead{$\delta_1$[dms]} & \colhead{$\alpha_2$[hms]} & \colhead{$\delta_2$[dms]} & \multicolumn{2}{c}{\textit{u}} & \multicolumn{2}{c}{\textit{g}} & \multicolumn{2}{c}{\textit{r}} & \multicolumn{2}{c}{\textit{i}} & \multicolumn{2}{c}{\textit{z}} }

\startdata
GBL0006+0154&0001-00341-2&   1237678596480565480&00 06 27.88& 01 55 16.0&00 06 23.29& 01 54 19.1&12.45&24.64&10.95&20.40&10.49&18.89&10.34&17.06&10.29&16.06\\
GBL0107+0057&0019-01107-1&   1237663785278570536&01 07 52.88&-00 57 40.7&01 07 53.11&-00 57 23.4&13.02&22.29&11.74&19.18&11.38&17.66&11.25&16.45&11.22&15.80\\
GBL0301+0500&0061-01165-1&   1237667227691384860&03 01 17.04& 05 00 02.7&03 01 19.86& 05 00 19.2&13.38&21.72&11.76&19.28&11.22&17.82&11.03&16.38&10.98&15.57\\
GBL0518+0104&0100-01294-1&   1237646588246425669&05 18 02.65&-01 03 53.4&05 18 03.15&-01 04 07.6&13.72&23.63&12.44&21.27&12.07&19.82&11.94&18.37&11.92&17.60\\
GBL0832+0055&0210-00761-1&   1237650797285867769&08 32 48.69& 00 55 53.1&08 32 51.42& 00 55 37.9&12.90&19.10&11.63&16.50&11.26&15.06&11.11&14.27&11.12&13.84\\
GBL0947+0016&0236-00406-1&   1237648721753079964&09 47 37.59&-00 15 07.2&09 47 33.29&-00 17 30.1&11.90&21.51&10.54&19.17&10.14&17.75& 9.99&16.36& 9.94&15.65\\
GBL1021+0003&0246-01091-1&   1237654670274789707&10 21 26.77&-00 04 09.3&10 21 27.69&-00 03 42.2&12.76&23.20&11.23&20.82&10.74&19.20&10.60&17.49&10.56&16.52\\
GBL1022+0501&0252-01639-1&   1237654606397505723&10 22 27.96&-05 01 47.2&10 22 21.60&-05 00 47.9&13.34&21.55&11.82&18.70&11.33&17.28&11.19&16.15&11.15&15.56\\
GBL1052+0028&0255-00257-1&   1237674650994999634&10 52 07.75&-00 29 35.4&10 52 10.34&-00 27 42.7&12.91&24.10&11.06&21.87&10.45&20.33&10.24&18.59&10.16&17.62\\
GBL1110+0224&0263-00181-1&   1237651754004906188&11 10 21.66&-02 25 09.9&11 10 22.92&-02 24 40.0&12.27&22.70&10.75&20.02&10.26&18.53&10.12&17.13&10.08&16.39\\
GBL1127+0127&0264-00873-1&   1237654028718047402&11 27 19.47& 01 27 23.8&11 27 17.69& 01 27 31.0&12.71&22.70&11.18&18.97&10.69&17.54&10.55&16.19&10.51&15.45\\
GBL1102+0307&0265-01094-1&   1237654030862909537&11 02 42.01&-03 07 30.8&11 02 44.13&-03 07 46.4&11.77&21.90&10.40&19.68&10.01&18.25& 9.85&16.81& 9.81&15.99\\
GBL1156+0024&0273-00397-1&   1237648721767235677&11 56 17.99&-00 24 46.2&11 56 21.84&-00 24 48.8&13.49&21.03&11.85&18.42&11.33&17.12&11.16&15.87&11.12&15.19\\
GBL1158+0448&0276-00104-1&   1237671140409868430&11 58 47.51&-04 49 53.3&11 58 43.03&-04 47 36.0& 9.72&21.87& 8.60&18.83& 8.32&17.47& 8.24&16.01& 8.25&15.22\\
GBL1358+0535&0315-00096-1&   1237661970122539182&13 58 23.97&-05 36 08.5&13 58 31.72&-05 35 25.1&10.43&20.88& 9.15&18.21& 8.78&16.69& 8.66&15.21& 8.63&14.36\\
GBL1538+0150&0351-00948-1&   1237651736854593630&15 38 21.34&-01 50 46.1&15 38 23.03&-01 50 50.5&14.15&22.18&12.46&19.41&11.86&18.00&11.65&16.56&11.59&15.78\\
GBL2133+0315&0546-01837-1&   1237678776851562540&21 33 54.15&-03 16 01.0&21 33 56.39&-03 15 54.9&11.32&23.32&10.01&19.67& 9.56&18.11& 9.44&16.49& 9.41&15.59\\
GBL2219+0546&0565-01839-1&   1237669761722941521&22 19 19.00&-05 45 29.2&22 19 29.90&-05 46 44.8&11.35&23.10&10.15&21.73& 9.91&20.21& 9.88&18.36& 9.89&17.38\\
GBL2341+0601&0592-01401-1&   1237672763912487004&23 41 26.68&-06 02 16.4&23 41 21.79&-06 01 37.4&12.99&21.45&11.47&18.92&10.98&17.47&10.84&16.14&10.79&15.41\\
GBL0024+1053&0599-00489-1&   1237678907858944027&00 24 39.12&-10 53 14.2&00 24 40.65&-10 53 27.9&12.14&26.06&10.87&20.77&10.49&19.25&10.35&17.56&10.36&16.66
\enddata

\tablecomments{*: This is only a part of the full table to be released online. The GAMBLES ID is created using the mean position of the pair.}
\end{deluxetable}
\end{turnpage}
\clearpage


\begin{deluxetable}{ccccccccccccc}

\tabletypesize{\tiny}
\tablewidth{0pt}
\tablecaption{Astrometric Information for GAMBLES Long-Lived Wide Binaries \label{tb:cand_ast}}
\tablehead{\colhead{ID} &  \multicolumn{4}{c}{Proper Motions [mas/yr$^{-1}$]} & \multicolumn{3}{c}{Separation} & \multicolumn{2}{c}{Distance [pc]} & \colhead{V$_{5}$} \\
 & \colhead{$\mu_{\alpha1}$} & \colhead{$\mu_{\delta1}$} &\colhead{$\mu_{\alpha2}$} & \colhead{$\mu_{\delta2}$} & \colhead{[pc]} & \colhead{[AU]} & \colhead{$\theta$ [\arcsec]} & \colhead{D$_1$} & \colhead{D$_2$} & }

\startdata
GBL0006+0154&  32.57$\pm$ 1.89& -59.56$\pm$ 0.78&  29.32$\pm$ 3.59& -57.55$\pm$ 3.59&0.06& 12979.1&   89.34&121&108&0.001\\
GBL0107+0057&  41.85$\pm$ 3.97&   3.15$\pm$ 1.15&  40.48$\pm$ 2.92&  -0.98$\pm$ 2.92&0.03&  5360.6&   17.69&235&245&0.001\\
GBL0301+0500&  44.07$\pm$ 2.75&  13.78$\pm$ 1.45&  40.00$\pm$ 3.03&  16.82$\pm$ 3.03&0.05&  9532.3&   45.22&178&156&0.001\\
GBL0518+0104&  31.68$\pm$ 1.29& -24.66$\pm$ 1.20&  37.45$\pm$ 5.38& -21.59$\pm$ 5.38&0.04&  9061.9&   16.05&487&408&0.001\\
GBL0832+0055& -28.63$\pm$ 3.43& -38.48$\pm$ 2.12& -31.98$\pm$ 2.84& -38.97$\pm$ 2.84&0.05& 10890.7&   43.72&203&191&0.001\\
GBL0947+0016& -42.07$\pm$ 1.46&  -2.18$\pm$ 0.90& -43.85$\pm$ 3.40&   2.47$\pm$ 3.40&0.20& 40560.7&  156.80&222&188&0.001\\
GBL1021+0003& -53.28$\pm$ 1.81&  23.67$\pm$ 2.05& -49.59$\pm$ 4.72&  26.08$\pm$ 4.72&0.03&  6182.5&   30.42&168&154&0.001\\
GBL1022+0501&  13.94$\pm$ 3.14& -44.77$\pm$ 1.56&  18.76$\pm$ 3.16& -46.11$\pm$ 3.16&0.17& 35540.4&  112.02&249&253&0.001\\
GBL1052+0028& -24.51$\pm$ 2.05& -44.71$\pm$ 1.41& -27.45$\pm$ 4.72& -43.40$\pm$ 4.72&0.17& 35976.9&  119.12&227&252&0.001\\
GBL1110+0224& -52.86$\pm$ 1.61&  -1.35$\pm$ 0.72& -55.84$\pm$ 3.55&  -0.45$\pm$ 3.55&0.05& 10094.6&   35.34&211&242&0.001\\
GBL1127+0127&  14.56$\pm$ 1.74& -51.20$\pm$ 0.82&   9.49$\pm$ 3.32& -51.23$\pm$ 3.32&0.03&  5882.8&   27.61&173&164&0.001\\
GBL1102+0307& -49.23$\pm$ 1.57&  -0.79$\pm$ 0.85& -46.83$\pm$ 3.40&   2.67$\pm$ 3.40&0.04&  8726.5&   35.39&208&182&0.001\\
GBL1156+0024& -54.31$\pm$ 2.07& -52.37$\pm$ 0.86& -60.68$\pm$ 6.06& -47.17$\pm$ 6.06&0.06& 12322.1&   57.73&166&172&0.001\\
GBL1158+0448& -36.73$\pm$ 0.05&  13.64$\pm$ 0.03& -38.49$\pm$ 2.93&  14.49$\pm$ 2.93&0.12& 25347.6&  152.75&136&127&0.001\\
GBL1358+0535& -64.67$\pm$ 0.76& -19.49$\pm$ 0.52& -65.24$\pm$ 2.64& -20.37$\pm$ 2.64&0.06& 12858.6&  123.55& 85& 79&0.001\\
GBL1538+0150& -41.58$\pm$ 1.32& -26.45$\pm$ 0.77& -40.32$\pm$ 3.18& -25.15$\pm$ 3.18&0.03&  5729.9&   25.62&182&172&0.001\\
GBL2133+0315&  87.32$\pm$ 1.24&  43.38$\pm$ 1.14&  86.95$\pm$ 3.44&  38.27$\pm$ 3.44&0.02&  4988.0&   34.01&114&118&0.001\\
GBL2219+0546& -20.90$\pm$ 1.21& -32.26$\pm$ 0.63& -21.24$\pm$ 4.68& -34.71$\pm$ 4.68&0.23& 47720.6&  179.41&215&206&0.003\\
GBL2341+0601&  95.46$\pm$ 3.22& -36.43$\pm$ 1.22&  93.17$\pm$ 2.77& -34.17$\pm$ 2.77&0.09& 18636.7&   82.77&183&173&0.001\\
GBL0024+1053&  62.51$\pm$ 1.85& -14.36$\pm$ 0.86&  59.07$\pm$ 3.93& -12.25$\pm$ 3.93&0.03&  6323.9&   26.36&196&184&0.001
\enddata

\tablecomments{*: This is only a part of the full table to be released online.}
\end{deluxetable}

\clearpage

\begin{deluxetable}{ccccccccc}

\tabletypesize{\tiny}
\tablewidth{0pt}
\tablecaption{Calculated Quantities for GAMBLES Long-Lived Wide Binaries \label{tb:cand_cal}}
\tablehead{\colhead{ID} &  \multicolumn{2}{c}{Spectral Type} & \multicolumn{2}{c}{Mass [M$_{\odot}$]} & \multicolumn{2}{c}{Reduced Proper Motion} & \colhead{Binding Energy} & \colhead{Dissipation Lifetime} \\
 & \colhead{SpT$_1$} & \colhead{SpT$_2$} & \colhead{M$_1$} & \colhead{M$_2$} &\colhead{H$_{g1}$} & \colhead{H$_{g2}$} & \colhead{[log$_{10}$(ergs)]} & \colhead{[Gyr]} }

\startdata
GBL0006+0154&G3&M4&1.0&0.2&10.11&19.37&41.38& 22.87\\
GBL0107+0057&G0&M2&1.1&0.4& 9.78&17.13&42.10& 66.70\\
GBL0301+0500&G3&M3&1.0&0.3& 9.95&17.02&41.73& 34.01\\
GBL0518+0104&G1&M3&1.1&0.3&10.46&19.05&41.77& 37.16\\
GBL0832+0055&G0&M0&1.1&0.6&10.03&14.87&41.99& 37.55\\
GBL0947+0016&G3&M3&1.0&0.3& 8.64&17.01&41.15&  8.25\\
GBL1021+0003&G2&M4&1.0&0.2&10.06&19.42&41.75& 49.33\\
GBL1022+0501&G0&M2&1.1&0.4&10.08&17.12&41.32& 10.28\\
GBL1052+0028&G7&M4&0.9&0.2& 9.60&20.28&40.93&  7.76\\
GBL1110+0224&G2&M3&1.0&0.3& 9.36&18.64&41.71& 32.55\\
GBL1127+0127&G2&M3&1.0&0.3& 9.81&17.46&41.96& 56.31\\
GBL1102+0307&G3&M3&1.0&0.3& 8.85&17.86&41.73& 36.60\\
GBL1156+0024&G4&M2&1.0&0.3&11.24&17.78&41.69& 27.18\\
GBL1158+0448&F6&M3&1.1&0.3& 6.56&16.83&41.31& 13.74\\
GBL1358+0535&G1&M3&1.1&0.2& 8.29&17.28&41.55& 25.29\\
GBL1538+0150&G6&M3&0.9&0.3&10.92&17.60&41.90& 53.24\\
GBL2133+0315&G0&M4&1.1&0.2& 9.91&19.35&41.91& 64.50\\
GBL2219+0546&F4&M4&1.2&0.2& 8.07&19.45&40.90&  7.18\\
GBL2341+0601&G1&M3&1.0&0.3&11.44&18.58&41.50& 18.30\\
GBL0024+1053&F8&M4&1.1&0.2& 9.76&19.29&41.82& 52.33
\enddata

\tablecomments{*: This is only a part of the full table to be released online.}
\end{deluxetable}

\end{document}